\documentclass[journal]{IEEEtran}

\ifCLASSINFOpdf
\else
\fi

\usepackage{graphicx}
\usepackage{url}
\usepackage{caption}
\usepackage{cite}
\usepackage{subcaption}
\usepackage{multirow}
\usepackage{bmpsize}
\usepackage{blkarray,kbordermatrix,subfig,mathtools,amsmath,epsfig,amssymb,enumerate,graphicx,bbm,paralist,multirow,bm,booktabs,hhline,tabularx}
\usepackage{xcolor}
\usepackage[english]{babel}
\usepackage[linesnumbered,ruled,vlined]{algorithm2e}
\usepackage{algpseudocode}
\usepackage{verbatim}
\usepackage{epstopdf}
\usepackage{amsthm}

\newtheorem{proposition}{Proposition}
\newtheorem*{remark}{Remark}

\usepackage{pgf}
\usepackage{tikz}
\usepackage[utf8]{inputenc}
\usetikzlibrary{arrows,automata}
\usetikzlibrary{positioning}

\tikzset{
    state/.style={
           rectangle,
           rounded corners,
           draw=black, very thick,
           minimum height=2em,
           inner sep=2pt,
           text centered,
           },
}

\begin{document}

\title{Evaluation of Decentralized Event-Triggered Control Strategies for Cyber-Physical Systems}
\author{Sokratis~Kartakis,~Anqi~Fu,~Manuel~Mazo,~Jr.~\IEEEmembership{Member,~IEEE},~and~Julie~A.~McCann~\IEEEmembership{Member,~IEEE}
\thanks{Sokratis Kartakis~(\textit{Corresponding author}) and Julie A. McCann are with the Department of Computing, Imperial College London, London SW72AZ, UK
(e-mail: \{s.kartakis13,~j.mccann\}@imperial.ac.uk).}
\thanks{Anqi Fu and Manuel Mazo Jr. are with the Delft Center for Systems and Control, Delft University of Technology, Delft 2628CD, The Netherlands
(e-mail: \{a.fu-1, m.mazo\}@tudelft.nl).}}

\maketitle

\begin{abstract}
Energy constraint long-range wireless sensor/ actuator based solutions are theoretically the perfect choice to support the next generation of city-scale cyber-physical systems. Traditional systems adopt periodic control which increases network congestion and actuations while burdens the energy consumption. Recent control theory studies overcome these problems by introducing aperiodic strategies, such as event trigger control.
In spite of the potential savings, these strategies assume actuator continuous listening while ignoring the sensing energy costs. In this paper, we fill this gap, by enabling sensing and actuator listening duty-cycling and proposing two innovative MAC protocols for three decentralized event trigger control approaches. A laboratory experimental testbed, which emulates a smart water network, was modelled and extended to evaluate the impact of system parameters and the performance of each approach. Experimental results reveal the predominance of the decentralized event-triggered control against the classic periodic control either in terms of communication or actuation by promising significant system lifetime extension.

\end{abstract}


\ifCLASSOPTIONpeerreview
\begin{center} \bfseries EDICS Category: 3-BBND \end{center}
\fi
\IEEEpeerreviewmaketitle

\section{Introduction}
\label{section:Introduction}

Over the last decade, there has been a growing trend in industry to transform large-scale manual control and monitoring systems, such as electrical grids and water networks, into fully automatic Cyber-Physical Systems (CPS). The aim of this transformation is the improvement of quality of service  and reduction of maintenance cost.  In order to achieve these goals, plants and physical environments have been augmented with sensor and actuator nodes which enable monitoring and control by communicating wirelessly and periodically to data centres or local base stations. However, these periodic dynamic control implementations introduce communication and energy consumption overheads.

In large scale CPS, sensor and actuator nodes are usually energy constraint and installed in harsh environments. For example, in smart water network more than 97\% of sensing and actuation assets are located underground and powered by batteries \cite{swig2014}. To transmit the required information through long-range (several kilometres) wireless communications, high transmission power is required that leads to fast battery depletion. In addition, the periodic sensing, transmission, and actuation, regardless the state of the plant, decreases network bandwidth and increases actuations and consequently the energy consumption. Recent control theory studies propose to solve these problems by introducing aperiodic strategies, such as Event-Triggered Control (ETC) strategies, e.g. \cite{tabuada2007event, wang2011event, heemels2013periodic, mazo2011decentralizedtac, mazo2014asynchronous, khashooei2015event} in which the sensors and actuators communicate only when necessary.  %which guarantee stability and performance while minimizing energy consumption and network bandwidth usage. In Event Trigger Control (ETC) systems, the sensors and actuators communicate only when the stability or a predefined control performance are about to diminish.

In spite of the potential of significant savings, ETC techniques have only been partially examined and implemented on real systems, i.e. \cite{sandee2006analysis, HeeBor_ECC16a, EekRao_EBCCSP16a, lehmann2011extension, sigurani2015experimental, altaf2011wireless, lemmon2010event, trimpe2011experimental, orihuela2016suboptimal, blevins2015event, boisseau2015attitude, araujo2014system}. In \cite{EekRao_EBCCSP16a}, the authors propose a system based on the Diddyborg robot and examine the strategy presented in \cite{khashooei2015event}. However, this system is first-order and therefore unable to test complex event-triggered strategies. In \cite{araujo2014system}, an experimental evaluation was made for time-triggered control and event-triggered control from \cite{tabuada2007event}. However, this work requires state monitoring continuously to check event conditions. Additionally, the results in \cite{tabuada2007event} can only be used for system with collocated sensors. To the best of our knowledge, there is no experiment that validates and compares different decentralized event-triggered mechanisms under the same conditions. %Furthermore,  other aperiodic strategy, i.e. self-triggered control \cite{mazo2010iss}, are still limited to physical experiments, e.g. \cite{santos2014adaptive, araujo2011self, velasco2010experimental, camacho2010self}.

ETC systems have been studied extensively in order to guarantee convergence of plants under reduced communication schemes. However, the design and implementation of a communication protocol, which fully exploits the ETC behavior and ensures optimal communication, is still missing \cite{miskowicz2015event}. State of the art ETC approaches that are focused on communication, i.e.  \cite{ blind2013time, cervin2008scheduling, ramesh2014stability, demirel2015trade, ramesh2011steady, blind2011analysis, vilgelm2016adaptive, obermaisser2015event, mazo2015decentralized},%, araujo2014system
 have been limited to simulate or analyze theoretically the impact of network states on system performance. CSMA \cite{cali2000dynamic}, TDMA \cite{miao2016fundamentals} and ALOHA \cite{abramson1970aloha} were the three communication protocols which have been used in the above approaches. Specifically, \cite{blind2013time, cervin2008scheduling} provide useful insights and comparison of all the above communication protocols. The authors in \cite{ramesh2014stability, demirel2015trade, ramesh2011steady} present a Markov model that captures the joint interactions of the event-triggering policy and a contention resolution mechanism over CSMA communication.  In \cite{blind2011analysis, vilgelm2016adaptive}, the ALOHA protocol, which has been applied in Long Term Evolution (LTE) Random Access (RA) procedure, was combined with ETC, with \cite{vilgelm2016adaptive} to introduce the impact of collisions into the system performance. TDMA-based communication protocols, i.e. Time Triggered Network-on-a-Chip (TTNoC), Time Triggered Controller Area Network (TTCAN), were analysed in \cite{ obermaisser2015event}, which discusses their application to ETC systems. The earliest practical work is \cite{araujo2014system}, continued in \cite{mazo2015decentralized}, by proposing the extension of the TDMA-based IEEE 802.15.4 MAC layer \cite{ieee8022006} which has been used in communication protocols for network control via wireless, i.e. WirelessHART \cite{wirelesshart2007}. However, the main drawbacks of this approach are the assumption that the actuator nodes listen continuously to network messages. Furthermore, none of the prior work has considered the cost of sensing.  For example, in our evaluation platform \cite{kartakis2015waterbox}, the sensing costs almost the half the energy consumption of communication. In industrial systems with more energy hungry sensors, e.g. laser based turbidity sensors for water quality, the energy cost may surpass communications.

%In this paper, we not only propose a dynamic and real time control system for large scale CPS, but also merge the event trigger control theory with real industrial implementations by introducing innovative communication schemes. The combination of our proposed communication schemes and ETC approaches constitutes an ETC framework for CPS. Specifically, the contributions of this paper are listed as follows:

Uniquely, in this paper, the proposed system duty cycle the sensing and actuator listening activities and at the same time enable decentralized ETC techniques and introducing innovative communication schemes. Specifically, the contributions of this paper are listed as follows:
\begin{itemize}
\item Practical combination of sensor duty-cycling with three decentralized aperiodic control approaches.

\item A new MAC protocol that facilitates decentralized synchronous ETC without the requirement of continuous actuator communication.

\item A novel flexible MAC protocol that can also accommodate two decentralized and asynchronous ETC approaches, communicating firstly  absolute states and alternatively relative states only.
%\item We propose the design and implementation of three innovative TDMA-based MAC protocols, which enable the application of ETC technologies to real industrial control scenarios. To the best of our knowledge, this is the first implementation of new MAC layers, and not modification of an existing one, which specifically designed to exploit the behaviour of network control systems and the needs of ETC specifically.

%\item We bridge the gap between the theory of Event Trigger Control (ETC) and the large scale implementations, by presenting analytically the procedure of system identification, hybrid controller design and ETC application to a real plant. Smart water networks have been used, as a proof of concept.
\end{itemize}

By using an extended version of the WaterBox testbed environment \cite{kartakis2015waterbox}, we provide experimental results from  Time-Triggered Control (TTC) and four different ETC techniques: Periodic centralized ETC (PETC) \cite{heemels2013periodic}, Periodic Synchronous Decentralized ETC (PSDETC) \cite{mazo2011decentralizedtac}, and Periodic Asynchronous Decentralized ETC (PADETC) by transmitting absolute or relative state \cite{fu2016peridic}. To the best of our knowledge, this is the first real deployment of most of the implemented ETC techniques to a real plant.

\section{Event-Trigger Control Techniques}
\label{section:ETC}

%In this section, we give a brief introduction of how to apply the Time-Triggered Control (TTC) strategy; Periodic centralized Event-Triggered Control (PETC) strategy from \cite{heemels2013periodic}; Periodic Synchronous Decentralized Event-Triggered Control (PSDETC) strategy from \cite{mazo2011decentralizedtac}; and Periodic Asynchronous Decentralized Event-Triggered Control (PADETC) strategy from \cite{fu2016peridic}.

We denote the positive real numbers by $\mathbb{R}^+$, the natural numbers including zero by $\mathbb{N}$. $|\cdot|$ denotes the Euclidean norm in the appropriate vector space, when applied to a matrix $|\cdot|$ denotes the $l_2$ induced matrix norm. A matrix $P\in \mathbb{R}^{n\times n}$ is said to be positive definite, denoted by $P\succ0$, whenever $x^{\mathrm{T}}Px>0$ for all $x\neq 0$, $x\in\mathbb{R}^n$. For the sake of brevity, we write symmetric matrices of the form $\begin{bmatrix}
A & B \\
B^{\mathrm{T}} & C \\
\end{bmatrix}$ as $\begin{bmatrix}
A & B \\
\star & C \\
\end{bmatrix}$.

\subsection{Periodic control}

Consider a linear time-invariant (LTI) plant and controller
\begin{equation}\label{eq:ltisystem}
\dot\xi(t)=A\xi(t)+Bv(t),
\end{equation}
\begin{equation}\label{eq:controller}
v(t)=K\xi(t),
\end{equation}
where $\xi(t)\in\mathbb{R}^n$ is the state vector and $v(t)\in\mathbb{R}^m$ is the input vector at time $t$. Assume $A+BK$ is Hurwitz, the system is completely observable, and each sensor can access only one of the system states.

A sample-and-hold mechanism is implemented for the controller (\ref{eq:controller}):
\begin{equation}\label{eq:sampleandholdcontroller}
v(t)=K\hat\xi(t),
\end{equation}
where
\begin{equation}\label{eq:samplestate}
\hat\xi(t):=\xi(t_b),t\in[t_b,t_{b+1}[.
\end{equation}
and $\{t_b\}_{b\in\mathbb{N}}$ is the sequence of the update time of the state. Representing the sample-and-hold effect as a measurement error, we have:
\begin{equation}\label{eq:error}
\varepsilon(t):=\hat\xi(t)-\xi(t).
\end{equation}

Define $T$ as the sample period. In a periodic time-triggered control strategy, $t_b$ is determined by
\begin{equation}\label{eq:samplesequence}
%t=\{t_b|b\in\mathbb{N},\,t_{b+1}-t_{b}=h\}.
\{t_b|t_b=bT,b\in\mathbb{N},T>0\}.
\end{equation}

\subsection{Periodic centralized event-triggered control}

In event-triggered control strategies, the control input update time is determined by some pre-designed conditions. These conditions are always a relation between system state and sample-and-hold error (\ref{eq:error}). Therefore, control executions happen only when necessary. However, the centralized event-triggered condition presented in \cite{tabuada2007event} requires the continuous monitoring and transmission of the current state to check the event conditions. If the state cannot be measured continuously, we can either compute a stricter event condition considering measurement delays; or apply the %periodic event-triggered control
PETC strategy from \cite{heemels2013periodic}, which combines periodic sampled-data control and event-triggered control:

Consider system (\ref{eq:ltisystem}), (\ref{eq:sampleandholdcontroller}), (\ref{eq:error}), and a sample sequence (\ref{eq:samplesequence}). At each sampling time $t_b$, the controller updates its state by
\begin{equation}\label{eq:periodicetcstateapproximation}
\hat\xi(t_b)=\left\{
\begin{aligned}
&\xi(t_b), \text{ when }\xi_p^{\mathrm{T}}(t_b)Q\xi_p(t_b)>0\\
&\hat\xi(t_{b-1}), \text{ when }\xi_p^{\mathrm{T}}(t_b)Q\xi_p(t_b)\leq0,\\
\end{aligned}\right.
\end{equation}
where $\xi_p(t)=\begin{bmatrix}
\xi^{\mathrm{T}}(t) & \hat\xi^{\mathrm{T}}(t)
\end{bmatrix}^{\mathrm{T}}$, $Q$ satisfies $
Q:=\begin{bmatrix}
    (1-\sigma)I & -I \\
    -I & I \\
  \end{bmatrix}$,
and $\sigma>0$.

For the system (\ref{eq:ltisystem}), (\ref{eq:sampleandholdcontroller}), (\ref{eq:error}), and (\ref{eq:samplesequence}), if $\exists c>0$ and $\rho>0$ such that for any initial condition $\xi(0)\in\mathbb{R}^{n}$, $\forall t\in\mathbb{R}^+$, $|\xi(t)|\leq ce^{-\rho t}|\xi(0)|$ is satisfied, then the system is said to be globally exponential stable, we call $\rho$ the decay rate \cite{sontag2008input}.

According to Corollary III.3 in \cite{heemels2013periodic}, given a decay rate $\rho>0$, if there exist a matrix $P\succ0$ and scalars $\mu_i\geq 0$, $i\in\{1,2\}$, such that
\begin{equation}\label{eq:petclmi}
\begin{bmatrix}
  e^{-2\rho T}P+(-1)^i\mu_iQ & J_i^{\mathrm{T}}e^{\bar{A}^{\mathrm{T}}T}P \\
  \star & P \\
\end{bmatrix}\succ0,\,i\in\{1,2\},
\end{equation}
where
\begin{equation*}
\bar{A}:=\begin{bmatrix}
           A & BK \\
           0 & 0 \\
         \end{bmatrix},\,
J_1:=\begin{bmatrix}
       I & 0 \\
       I & 0 \\
     \end{bmatrix},\,
J_2:=\begin{bmatrix}
       I & 0 \\
       0 & I \\
     \end{bmatrix},
\end{equation*}
then the system is globally exponential stable with a decay rate $\rho$.
%With consideration of transmission delays, we can simply solve (\ref{eq:petclmi}) with new sampling time $\tilde{h}:=h+\Delta$, where $\Delta<h$ is the maximum delay.

\subsection{Periodic synchronous decentralized event-triggered control}

The event-triggered strategies presented in (\ref{eq:periodicetcstateapproximation}) are centralized event-triggered strategies, since the event conditions require the whole vector of current state and current error. When the sensors are not co-located, decentralized event conditions are preferred. We introduce the %periodic synchronous decentralized event-triggered control
PSDETC strategy %, which is
 based on \cite{mazo2011decentralizedtac}.

For system (\ref{eq:ltisystem}), (\ref{eq:sampleandholdcontroller}), and (\ref{eq:error}), a decentralized event-triggering condition with periodic sampling (\ref{eq:samplesequence}) is given by:
\begin{equation}\label{eq:periodisetcstateapproximation}
\hat\xi(t_b)=\left\{
\begin{aligned}
&\xi(t_b), \text{ when }\exists i:\,\varepsilon_i^2(t_b)-\sigma\xi_i^2(t_b)>\theta_i\\
&\hat\xi(t_{b-1}), \text{ when }\forall i:\,\varepsilon_i^2(t_b)-\sigma\xi_i^2(t_b)\leq\theta_i,\\
\end{aligned}\right.
\end{equation}
where $\varepsilon_i(t)$ and $\xi_i(t)$ denote the $i$-th coordinates of $\varepsilon(t)$ and $\xi(t)$ respectively, and $\{\theta_i\}_{i\leq n}$ is a set of parameters. Define $\{t_k\}:=\{t_b|\exists i,\varepsilon_i^2(t_b)-\sigma\xi_i^2(t_b)>\theta_i\}$ the sequence of event times. The sequence $\{\theta_i\}_{i\leq n}$ is obtained solving at each event time $t_k$:
\begin{equation}\label{eq:setcthetacalculation}
\left\{
\begin{aligned}
\hat{G}_i(t_k+t_e)=&\hat{\varepsilon}_i^2(t_k+t_e)-\sigma \hat{\xi}_i^2(t_k+t_e)- \theta_i(k)\\
\hat{G}_i(t_k+t_e)=&\hat{G}_j(t_k+t_e),\,\forall i,j \in \{1,2,\cdots,n\}\\
\sum_{i=1}^n\theta_i(k)=&0,
\end{aligned}
\right.
\end{equation}
where for $t\in[t_k,t_{k+1}[$
\begin{equation*}
\begin{aligned}
\hat{\xi}_i(t)=&\xi_i(t_k)+\dot{\xi}_i(t_k)(t-t_k)+\frac{1}{2}\ddot{\xi}_i(t_k)(t-t_k)^2+\ldots\\
&+\frac{1}{q!}\xi^{(q)}_i(t_k)(t-t_k)^q\\
\hat{\varepsilon}_i(t)=&0-\dot{\xi}_i(t_k)(t-t_k)-\frac{1}{2}\ddot{\xi}_i(t_k)(t-t_k)^2-\ldots\\
&-\frac{1}{q!}\xi^{(q)}_i(t_k)(t-t_k)^q.
\end{aligned}
\end{equation*}
The map $t_e:\mathbb{N}\rightarrow\mathbb{R}^+$ can be set to either $t_e(k)=T$ or $t_e(k)=t_k-t_{k-1}$. We apply Algorithm 1 in \cite{mazo2011decentralizedtac} to determine $t_e(k)$ in the experiments.

Thus, with the current $\theta_i(k)$ being calculated and transmitted from the controller to each sensor node, the sensor node can locally determine the occurrence of local events. When there is an event, the corresponding sensor node notifies the controller, and then the controller requests fresh measurements from all sensors to compute and update the control input.

\begin{proposition}\label{lemma}
For system (\ref{eq:ltisystem}), (\ref{eq:sampleandholdcontroller}), (\ref{eq:error}), and (\ref{eq:periodisetcstateapproximation}), given a decay rate $\rho>0$, if there exist a matrix $P\succ 0$ and scalars $\mu_1,\,\mu_2,\,\mu_3\geq0$, such that
\begin{equation}\label{eq:setclmi}\left\{
\begin{aligned}
\begin{bmatrix}
  e^{-2\rho T}P-\mu_1Q & J_1^{\mathrm{T}}e^{\bar{A}^{\mathrm{T}}T}P \\
  \star & P \\
\end{bmatrix}&\succ0\\
\begin{bmatrix}
  e^{-2\rho T}P+\mu_2Q & J_2^{\mathrm{T}}e^{\bar{A}^{\mathrm{T}}T}P \\
  \star & P \\
\end{bmatrix}&\succ0\\
\begin{bmatrix}
  e^{-2\rho T}P+\mu_3Q & J_1^{\mathrm{T}}e^{\bar{A}^{\mathrm{T}}T}P \\
  \star & P \\
\end{bmatrix}&\succ0,\\
\end{aligned}\right.
\end{equation}
hold, then the system is globally exponential stable with a decay rate $\rho$.
\end{proposition}
\begin{IEEEproof}
According to \cite{mazo2011decentralizedtac}, $\forall i:\,\varepsilon_i^2(t)-\sigma\xi_i^2(t)\leq\theta_i$ implies $\varepsilon^{\mathrm{T}}(t)\varepsilon(t)\leq\sigma\xi^{\mathrm{T}}(t)\xi(t)$, which is equivalent to  $\xi_p^{\mathrm{T}}(t_b)Q\xi_p(t_b)\leq0$. However, $\exists i:\,\varepsilon_i^2(t)-\sigma\xi_i^2(t)>\theta_i$ may indicate $\xi_p^{\mathrm{T}}(t_b)Q\xi_p(t_b)>0$ or $\xi_p^{\mathrm{T}}(t_b)Q\xi_p(t_b)\leq 0$. From the proof of Corollary III.3 in \cite{heemels2013periodic}, if the hypothesis in Proposition \ref{lemma} hold, by applying the S-procedure (see e.g. \cite{boyd1994linear}), one obtains
\begin{equation*}
\begin{aligned}
&x^{\mathrm{T}}Qx>0    \text{ holds, then } W(J_1x,0)\leq W(x,T),\\
&x^{\mathrm{T}}Qx\leq 0\text{ holds, then } W(J_2x,0)\leq W(x,T),\\
&x^{\mathrm{T}}Qx\leq 0\text{ holds, then } W(J_1x,0)\leq W(x,T).\\
\end{aligned}
\end{equation*}
in which $W(x,\tau)$ is a Lyapunov function defined as (18) in \cite{heemels2013periodic}. Therefore, $W$ does not increase during samplings. Together with the results from Theorem III.2 in \cite{heemels2013periodic} that, such a $P$ can guarantee $\frac{d}{dt}W\leq-2\rho W$ between samplings, the system is globally exponential stable with a decay rate $\rho$.
\end{IEEEproof}

\subsection{Periodic asynchronous decentralized event-triggered control}
\label{subsec:padetc}

A periodic asynchronous event-triggered control strategy is presented in \cite{fu2016peridic}. In this strategy, again the triggering condition is distributed to each sensor node. When there is an event, in contrast with PSDETC, only the corresponding sensor node data is used to update the controller. The updated control input is then calculated and transmitted to the actuators.

Consider system (\ref{eq:ltisystem}), (\ref{eq:sampleandholdcontroller}), and (\ref{eq:error}), the current sampled state in the controller is updated as:
\begin{equation}\label{eq:paetcsampleupdate}
\hat{\xi}_i(t_b)=\left\{\begin{aligned}
&q(\xi_i(t_b)),\,\text{if } i\in\mathcal{J}\\
&\hat{\xi}_i(t_{b-1}),\,\text{if } i\in\mathcal{J}_c,
\end{aligned}\right.
\end{equation}
where $q(\xi_i(t))$ denotes the quantized signal of $\xi_i(t)$, $\mathcal{J}$ is an index set: $\mathcal{J}\subseteq\{1,\cdots,n\}$ for $\xi(t)$, indicating the occurrence of events, $\mathcal{J}_c:=\{1,\cdots,n\}\setminus\mathcal{J}$.

Define $\Gamma_{\mathcal{J}}:=\text{diag}(\gamma^1_{\mathcal{J}} \cdots,\gamma_{\mathcal{J}}^{n})$. The element $\gamma_{\mathcal{J}}^l$, with $l\in\{1,\cdots,n\}$ is equal to 1, if $l\in\mathcal{J}$, and 0 otherwise. Furthermore, we use the notation $\Gamma_j=\Gamma_{\{j\}}$. The local event-triggering condition is:
\begin{equation}\label{eq:paetceventcondition}
\begin{aligned}
&i\in\mathcal{J}\text{ iff }\xi_p^{\mathrm{T}}(t_b)Q_i\xi_p(t_b)\geq\eta_i(t_b),\\
\end{aligned}
\end{equation}
where $Q_i:=\begin{bmatrix}
             \Gamma_i & -\Gamma_i \\
             -\Gamma_i & \Gamma_i \\
           \end{bmatrix}
$, $\eta_i(t):=\omega_i^2\eta^2(t)$ is a local threshold, $\omega$ is a pre-designed distributed parameter satisfying $|\omega|=1$, $\eta(t)$ is a global threshold, determined by
\begin{equation}\label{eq:paetcthresholdupdate}
\begin{aligned}
\eta(t_b^+)=
\left\{
\begin{aligned}
&\mu\eta(t_b),\text{ if }|\hat\xi(t_b^+)|\leq\varrho\eta(t_b) \wedge\eta(t_b)>\mu^{-1}\eta_{\min},\\
&\eta_{\min},\text{ if }|\hat\xi(t_b^+)|\leq\varrho\eta(t_b)\wedge\eta(t_b) \leq\mu^{-1}\eta_{\min},\\
&\mu^{-1}\eta(t_b),\text{ if }|\hat\xi(t_b^+)|\geq\mu^{-1}\varrho\eta(t_b),\\
&\eta(t_b),\text{ otherwise},
\end{aligned}\right.
\end{aligned}
\end{equation}
where $\eta_{\min}> 0$ is a pre-specified minimum threshold, and $\mu\in]0,1[$ is a pre-designed parameter.

Consider the Hamiltonian matrix:
\begin{equation}\label{eq:H}
\begin{aligned}
H:=\begin{bmatrix}
     H_{11} & H_{12} \\
     H_{21} & H_{22} \\
   \end{bmatrix},
\end{aligned}
\end{equation}
where $H_{11}:=\bar{A}+\rho I,\,H_{12}:=0,\,H_{21}:=-(\gamma^2I-I)^{-1},\,H_{22}:=-(\bar{A}+\rho I)^{\mathrm{T}}$, for some $\gamma>1$, and some given $\rho>0$. And introduce the matrix exponential
\begin{equation}\label{eq:ftau}
\begin{aligned}
F(\tau):=e^{-H\tau}=\begin{bmatrix}
                      F_{11}(\tau) & F_{12}(\tau) \\
                      F_{21}(\tau) & F_{22}(\tau) \\
                    \end{bmatrix}.
\end{aligned}
\end{equation}
Define the matrix $\bar{S}$ satisfying $\bar{S}\bar{S}^{\mathrm{T}}:=-F_{11}^{-1}(T)F_{12}(T)$.

According to Theorem IV.4 in \cite{fu2016peridic}, consider the system (\ref{eq:ltisystem}), (\ref{eq:sampleandholdcontroller}), (\ref{eq:error}), (\ref{eq:paetcsampleupdate}), and (\ref{eq:paetcthresholdupdate}), given a decay rate $\rho>0$. Assume $F_{11}(\tau)$ is invertible $\forall \tau\in[0,T]$. If there exist matrix $P\succ0$, scalars $\varrho>0$, $\beta_1>0$, $\beta_2>0$ and $\epsilon_{\mathcal{J}_i}>0$, $\mathcal{J}\subseteq\{1,\cdots,n\}$, $i\in\{1,\cdots,n\}$ such that the bilinear matrix inequality
\begin{equation}\label{eq:paetclmi}
\begin{aligned}
\begin{bmatrix}
  \beta_2 I & F_{11}^{-\mathrm{T}}(T)P\bar{S} & \tilde{F} & -\beta_2 J_{\mathcal{J}} & 0 \\
  \star & I-\bar{S}^{\mathrm{T}}P\bar{S} & 0 & 0 & 0 \\
  \star & \star & \tilde{F} & 0 & 0 \\
  \star & \star & \star & P+\tilde{H}_1 & 0 \\
  \star & \star & \star & \star & \tilde{H}_2 \\
\end{bmatrix}\succ 0,
\end{aligned}
\end{equation}
holds, where
\begin{equation*}
\begin{aligned}
&\tilde{F}:=F_{11}^{-\mathrm{T}}(T)P F_{11}^{-1}(T)+F_{21}(T)F_{11}^{-1}(T)\\
&\tilde{H}_1:=-\beta_1I+\beta_2J_{\mathcal{J}}^{\mathrm{T}}J_{\mathcal{J}}-\sum_{i\in\mathcal{J}} \epsilon_{\mathcal{J}_i}Q_i+\sum_{i\in\mathcal{J}_c}\epsilon_{\mathcal{J}_i}Q_i\\ &\tilde{H}_2:=\beta_1\varrho^2I-\beta_2 \bar\Delta_{\mathcal{J}}^{\mathrm{T}}\bar\Delta_{\mathcal{J}}+\sum_{i\in\mathcal{J}} \epsilon_{\mathcal{J}_i}\Theta^{\mathrm{T}}\Gamma_i\Theta \\ &-\sum_{i\in\mathcal{J}_c}\epsilon_{\mathcal{J}_i}\Theta^{\mathrm{T}}\Gamma_i\Theta\\
&\bar\Delta_{\mathcal{J}}:=\begin{bmatrix}
0 \\
\Gamma_{\mathcal{J}}\Theta \\
\end{bmatrix},\,J_{\mathcal{J}}=\begin{bmatrix}
                   I & 0 \\
                   \Gamma_{\mathcal{J}} & I-\Gamma_{\mathcal{J}} \\
                 \end{bmatrix}\\
&\Theta=\begin{bmatrix}
\omega_1 & \cdots & \omega_n
\end{bmatrix}^{\mathrm{T}},\\
\end{aligned}
\end{equation*}
then $\mathcal{A}:=\{x||x|\leq\bar\varrho\eta_{\min}\}$
is a globally pre-asymptotically stable set for the system (see e.g. \cite{goebel2012hybrid}), where $\bar\varrho:=\max\{|J_{\mathcal{J}}|\varrho+|\bar\Delta_{\mathcal{J}}|,\forall\mathcal{J}\}$.

In \cite{fu2016peridic}, the update of the signals $\hat{\xi}_i(t_b)$ is given by:
\begin{equation*}
\begin{aligned}
&\hat{\xi}_i(t_b)=\hat{\xi}_i(t_{b-1})+\text{sign}(\hat{\xi}_i(t_{b-1})-\xi_i(t_b))m_i(t_b)\sqrt{\eta_i(t_b)},\\
\end{aligned}
\end{equation*}
in which $m_i(t_b):=\lfloor\frac{|\hat{\xi}_i(t_{b-1})-\xi_i(t_b)|}{\sqrt{\eta_i(t_b)}}\rfloor,\,i\in\{1,\cdots,n\}$. Thus, in practise one only needs to send $\text{sign}(\hat{\xi}_i(t_{b-1})-\xi_i(t_b))$ and $m_i(t_b)$ from sensor to controller. We call PAETC with this update mechanism as PADETCrel, since we transmit the relative value (increment and its sign). A dynamic quantizer can be applied with maximum quantization error $\sqrt{\eta_i(t_b)}$ for each sensor. However, since the static quantizers we install in our experimental setting have quantization error neglectable compared with the noise, and to compare with the rest of strategies fairly, we can instead update signals $\hat{\xi}_i(t_b)$ by $\hat{\xi}_i(t_b)=\xi_i(t_b)$. We call PAETC with this update mechanism as PADETCabs, since we transmit the absolute value.

\section{Incorporating ETC with the MAC Layer}
\label{section:MAC}

In this section, we present the design and implementation of three innovative TDMA-based MAC protocols which enable the deployment of TTC, PETC, PSDETC, and PADETC approaches accordingly: Control-TDMA (C-TDMA), Synchronous Decentralized-CTDMA (SDC-TDMA), and Asynchronous Decentralized-CTDMA (ADC-TDMA). The main benefits of these ETC-specific MAC protocols are: the optimization of communications by fully exploiting the behaviour and needs of ETC; the minimization of actuator node listening through duty cycling; and the off-load of the local controller node (base station) by minimizing the interaction with sensor and actuator nodes. %Additionally, the design flexibility enables the protocols to be applicable to other communication technologies, e.g. FTDMA-based Weightless communication \cite{weightless2012}, which can be extended to support city-scale control systems that consist of millions of sensor/actuator nodes.
For the proposed TDMA-based communication schemes, we assume a CPS network architecture, %in Figure \ref{fig:netarchitecture},
in which the sensor/actuator nodes communicate with a based station and retrieve acknowledgement messages per transmission. Further, the controller which computes the control signals is executed in the base station. In this paper, the terms controller and base station is used interchangeably.

\subsection{Simplistic TDMA Protocol}
\begin {figure}[!ht]
\centering
\includegraphics[width=0.6\linewidth]{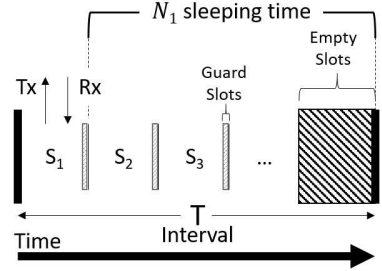}
\caption{Simplistic TDMA MAC Protocol.}
\label{fig:figtdma}
\end {figure}

% \begin{figure*}
%     \centering
%     \begin{subfigure}[b]{0.28\textwidth}
%         \includegraphics[width=\textwidth]{Figures/tdma.eps}
%         \caption{Simplistic TDMA MAC Protocol.}
%         \label{fig:figtdma}
%     \end{subfigure}
%     \centering
%     \begin{subfigure}[b]{0.52\textwidth}
%         \includegraphics[width=\textwidth]{Figures/ctdma.eps}
%         \caption{C-TDMA Protocol for TTC and PETC approaches.}
%         \label{fig:figctdma}
%     \end{subfigure}
%     \caption{Extending TDMA protocol to support ETC techniques.}
%      \label{fig:figextdma}
% \end{figure*}

TDMA is a channel access method for shared medium networks, which allows several users to share the same frequency channel. Specifically, time is divided into intervals $T_i$ with length $T$, so-called super-frames. Each interval is split into smaller time slots $S_j$, with $\sum\limits_{j=1}^N S_j \leq T$, where $N$ is the number of sensor/actuator nodes which share the same channel\footnote{A super-frame can be divided into equal time slots that fully utilize the channel or to minimal slots which cover the application requirements and allow the communication to new nodes into the same channel. }. In each time slot, only one predefined sensor/actuator node $N_j$ can transmit ($T_x$) or receive ($R_x$) a burst of messages to and from a base station. Outside the timeslot $S_j$, $N_j$ sleeps or executes other tasks depending on the hardware infrastructure and the provided ability to duty cycle. To avoid time violations of time slot bounds due to $N_j$ possible clock drift, a guard slot forces the termination of communication before the end of each $S_j$. Figure \ref{fig:figtdma} illustrates the communication scheme of %the
a simplistic TDMA protocol.

%Due to synchronous operation, a TDMA-based protocol infrastructure can accommodate the development of ETC techniques in real plants. The main reason relies on the necessity of control systems to capture and transmit the state of the plant from all the sensor nodes concurrently (except the ADETC approach) in order to ensure control signal computation accuracy. Furthermore, a-priori knowledge of wake up scheduling prevents collisions due to reduced broadcast messages and local controller off-loading by serving specific sensor/actuator node per time slot. %For example, the local controller may avoid broadcast messages and bandwidth allocation for a long period of time, and serve the sensor/actuator nodes only whenever requests are being raised.

Due to synchronous operation, a TDMA-based protocol (e.g. \cite{wirelesshart2007}) can guarantee tight bounds on delays which are critical for network control systems. %(e.g. WirelessHART \cite{wirelesshart2007})
On the other hand, synchronizing sensor/actuator nodes is considered as the main drawback of TDMA-based systems. However, state of the art solutions, i.e. GPS clock synchronization technologies \cite{spectracom2008}, ensure typical accuracy better than 1 microsecond by consuming ultra low power.% This time synchronization technology has been tested widely, in term of robustness and performance, in real city-scale deployments, such as the Smart Water Network in \cite{hoskins2014infrasense}.

\subsection{C-TDMA and TTC \& PETC}
\begin {figure}[!ht]
\centering
\includegraphics[width=0.95\linewidth]{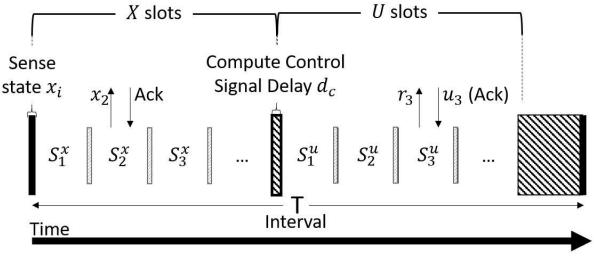}
\caption{CTDMA protocol.}
\label{fig:figctdma}
\end {figure}
%By having describe the architecture and benefits of a TDMA protocol, the next step is to describe our extended TDMA protocol which accommodate the communication under different ETC scenarios. The first protocol,
Control-TDMA (C-TDMA) is designed to enable TTC and PETC approaches (see Figure \ref{fig:figctdma}). In order to ensure the simultaneous state sampling%capturing
, in the beginning of every interval $T_i$ at time $t_i$, every node $N_j$ has to retrieve a state measurement $x_j$ from the available sensors. Then, the channel bandwidth is divided into two sets of time slots which are separated by a time delay:
\subsubsection{\textbf{Measurement Slots $\mathbf{S_j^x}$ (X-slots)}} Every sensor/actuator node $N_j$ transmits $x_j$ within the time slot $S_j^x$ to the controller. Within each time slot only one successful message is required. Thus, the size selection of $S_j^x$ is application specific and depends on the size of $x_j$ (e.g. 2 Bytes per sensor) and the number of re-transmissions to achieve high reliability based on the selected hardware.
\subsubsection{\textbf{Delay $\mathbf{d_c}$}} After receiving the complete sampled state by receiving $x_j,\forall j\in N$, a time delay is required to allow the computation of appropriate control signals $u_j$ for every sensor/actuator node $N_j$. The length of this delay depends on the controller infrastructure and the complexity of the control model.
\subsubsection{\textbf{Actuation Slots $\mathbf{S_j^u}$ (U-slots)}}The last set of time slots is related to the control message retrieval by the sensor/ actuator nodes $N_j$. In each time slot $ S_j^u $, node $N_j$ transmits a request $r_j$ for a control signal $u_j$ to the controller. Then, the $u_j$ is piggybacked to the acknowledgement message. The benefit of the $r_j$ request is two-fold: (a) off-loads the controller side and (b) reduces $N_j$ listening time. Otherwise, the controller has to transmit or broadcast $u_j$ continuously by blocking other tasks, while $N_j$ has to be active in receive mode during the full length of the $S_j^u$ slot until a successful control message retrieval. Further, this approach causes more energy savings for nodes with communication modules that consume the same amount of energy for transmission and listening, i.e. \cite{ti2016cc}. The length of $S_j^u$ depends on the size of  $u_j$ signals. %can be smaller than $S_j^x$, as $r_j$ requires the minimum supported message length (e.g. 1 Byte) and the $u_j$ has smaller or equal size than one measurement $x_j$.

Based on the above, the minimum interval size $T_{min}$ can be defined by the length of X-slots, U-slots and delay $d_c$, and the number of the nodes. Further, the $T_{min}$ can be considered as the maximum time delay of the system.

TTC and PETC are centralized approaches and are executed in the controller. In both cases, the system requires the transmission of the current state to the controller at every $T_i$ during the X-slots. Then, in the TTC technique, the controller replies back in the U-slots of every interval $T_i$ with a new $u_j$ control signal. On the other hand, in PETC, the controller evaluates the event condition, as has been described in Equation (\ref{eq:periodicetcstateapproximation}), and transmits the new $u_j$ only if there is a violation. This behaviour allows PETC to save energy due to actuation reduction.
%As described above, C-TDMA enables the centralized TTC and PETC by introducing the U-slots and scheduling the sensing and control signal computation appropriately.
%Additionally, C-TDMA reduces the communication and listening time by avoiding burst message transmissions and introducing $r_j$ signals.
%The only unavoidable drawback is the increase of one slot set in TDMA to two slot sets (X-slots and U-slots) in C-TDMA. This increase affects only the selection of the minimum interval size $T_{min}$.
%However, due to collision avoidance and optimal task synchronization (sensing and actuation) in C-TDMA, $T_{min}$ is the minimum amount of time that a system with another MAC layer requires to operate a close loop control scenario.
%C-TDMA constitutes the fundamental infrastructure for the following ADC-TDMA and SDC-TDMA.

\subsection{SDC-TDMA and PSDETC}
\begin{figure*}
    \centering
    \begin{subfigure}[b]{0.58\textwidth}
        \includegraphics[width=\textwidth]{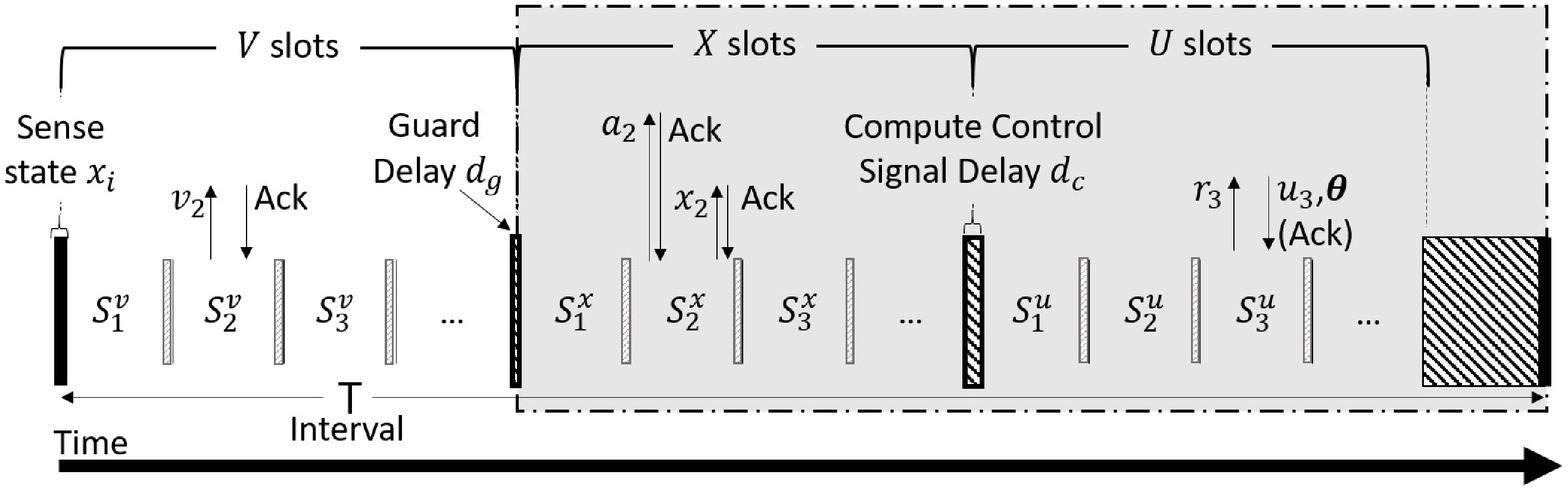}
        \caption{SDC-TDMA Protocol for PSDETC approach.}
        \label{fig:figsctdma}
    \end{subfigure}
    \quad
    \begin{subfigure}[b]{0.39\textwidth}
        \includegraphics[width=\textwidth]{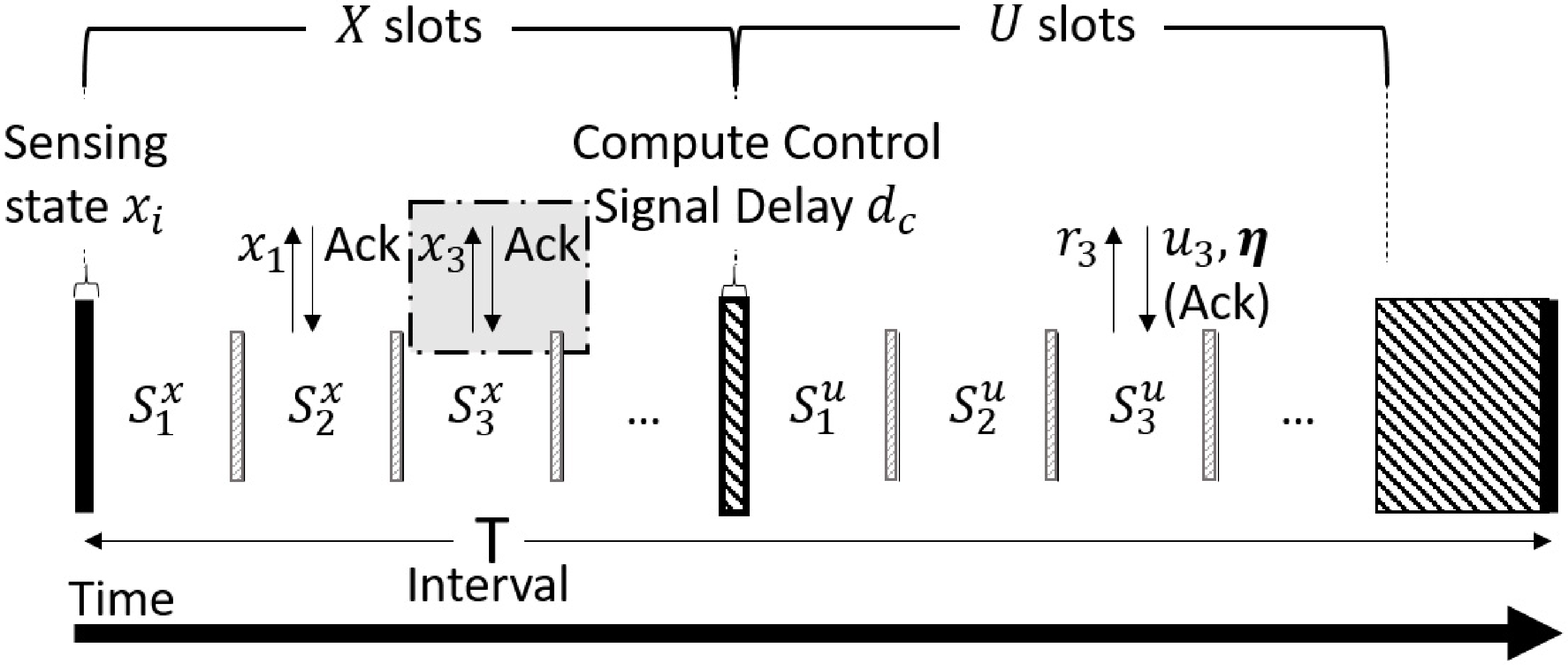}
        \caption{ADC-TDMA Protocol for PADETC approach.}
        \label{fig:figactdma}
    \end{subfigure}
    \caption{Communication schemes for decentralized ETC techniques.}
    \label{fig:figasctdma}
\end{figure*}
PSDETC is a distributed technique and each sensor node is responsible for the state transmission decision in every $T_i$. The computation of control signals $u_j$ and $\theta_j$ parameters require the \textbf{complete} knowledge of the system's state for the interval $T_i$ in which a threshold violation has happened. For example, consider a system with three nodes, $\{N_1, N_2, N_3\}$, in which only two of the nodes, i.e. $\{N_2,N_3\}$, observe threshold violations. The controller requires the state from all the three nodes to compute the control signals% accurately
. Using the same example, in a TDMA-based communication scheme in which each node is assigned to a pre-defined time slot $S_j$, node $N_1$ is precluded from transmitting its state by being unable to be informed about the threshold violations of $N_2$ and $N_3$.
%However, in a TDMA approach, each sensor node communicates with a local controller within a predefined time slot $S_j$. This limitation may lead to the lack of threshold violation notifications (e.g. $N_1$) from other nodes which communicate within future time slots (e.g. $N_2$ and $N_3$) of the same $T_i$.
To overcome these limitations, SDC-TDMA introduces a new set of time slots $S_j^v$, the V-slots (see Figure \ref{fig:figsctdma}).

\subsubsection{\textbf{Violations Slots $\mathbf{S_j^v}$ (V-slots)}} In the beginning of every $T_i$, each node retrieves a measurement and evaluates Equation (\ref{eq:periodisetcstateapproximation}%\ref{eq:sdetccondition}
). Then, the result of each threshold $v_j$ is transmitted by the corresponding node $N_j$ to the controller at time slot $S_j^v$. %The size of $S_j^v$ is the minimum available and depends on the selected hardware infrastructure (e.g. 1 Byte).
\subsubsection{\textbf{Measurement Slots $\mathbf{S_j^x}$ (X-slots)}}
%After the V-slots, the X-slots and U-slots follow (see Figure \ref{fig:figsctdma}).
In the beginning of every $S_j^x$ in X-slots, each node $N_j$ asks the controller, by sending an $a_j$ request, whether a threshold violation was observed in the V-slots. If no threshold violation occurred, the sensor node sleeps immediately until the next interval $T_{i+1}$ (gray box in Figure \ref{fig:figsctdma}). Otherwise, each node transmits each state $x_j$ to controller, wait for the delay $d_c$ and actuation slots, U-slots, follow.
\subsubsection{\textbf{Delay $\mathbf{d_c}$ \& Actuation Slots $\mathbf{S_j^u}$ (U-slots)}} Similar to the C-TDMA approach, after a delay $d_c$, each node requests the new control signal $u_j$ from the controller. The $u_j$ and the new threshold parameters $\theta _j$, which is being calculated based on Equation (\ref{eq:setcthetacalculation}), are being piggybacked into the acknowledgement messages of U-slots.

Based on the above, SDC-TDMA sacrifices channel availability and increases the minimum interval length, $T_{min}$ and consequently the system's maximum delay, by adapting V-slots into the TDMA scheme.  However, in the case of threshold violation absence, the communication is being minimized significantly by avoiding the transmission of state $x_j$ and the entire execution of U-slots.

\subsection{ADC-TDMA and PADETC}
Similar to PSDETC, the PADETC approach transfers the communication decision making from the controller down to the sensor/actuator nodes. Additionally, due to its asynchronous feature, this approach increases the communication savings by avoiding the state transmission $x_j$ from every node $N_j$ in every interval $T_i$. The only overhead in the communication is the $\eta _j$ update based on Equation (\ref{eq:paetcthresholdupdate}) and transmission to $N_j$. The value of $\eta _j$ is being piggybacked with the control signal $u_j$ in the acknowledgement message.

Specifically, the architecture of ADC-TDMA is similar to C-TDMA and consists of sensing task, X-slots, $d_c$ delay, and U-slot (see Figure \ref{fig:figactdma}). In a $S_j^x$ slot, the node $N_j$ evaluates the threshold of Equation (\ref{eq:paetceventcondition}). In the case of no threshold violation (i.e. gray box of Figure \ref{fig:figactdma}), the node $N_j$ skips the communication and returns to sleep mode until $S_j^u$. Otherwise, $N_j$ transmits to the controller: $x_j$ in PADETCabs or the increment $m_j$ in PADETCrel (see Section \ref{subsec:padetc}). Then, the controller computes the appropriate control signals and updates the local and global $\eta$ based on Equation (\ref{eq:paetcthresholdupdate}) by using \textbf{only} the available $x_j$ states. %For example, consider a system with three sensor/actuator nodes, at time interval $T_i$, the nodes $N_1$ and $N_2$ decide to transmit to the local controller the states $\xi_1(t_i)$ and $\xi_2(t_i)$ because of local threshold violations. Then the local controller computes the $u_j$ and $\eta _j$ based on $\xi_1(t_i)$, $\xi_2(t_i)$, and $\hat{\xi_3}(t_i)$.

In the U-slots, every node has to send a $r_j$ request message to the controller, in order to be notified about a possible threshold violation from another node. Therefore, the violation of at least one threshold causes the update of $u_j$ and $\eta _j$ to be sent to every actuator node. The values of $u_j$ and $\eta _j$ are piggybacked to the acknowledgement message.

%Section \ref{section:Evaluation} will provide further details about this communication trade-off by comparing the proposed TDMA-based schemes.

\section{Evaluation Platform: WaterBox}
\label{section:SystemOverview}

\begin {figure}[!ht]
\centering
\includegraphics[width=0.75\linewidth]{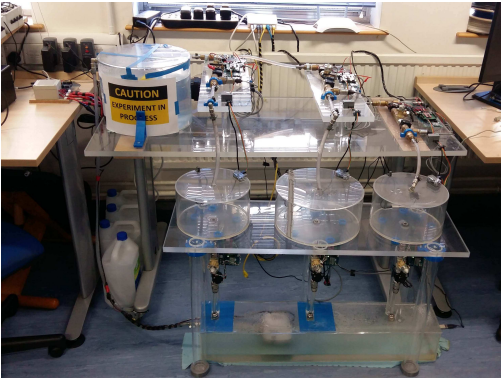}
\caption{WaterBox Testbed.}
\label{fig:waterboxtestbed}
\end {figure}

\begin{figure*}[ht!]
    \centering
    \begin{subfigure}[b]{0.37\textwidth}
        \includegraphics[width=\textwidth]{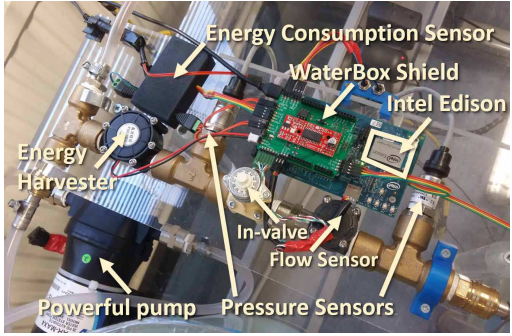}
        \caption{WaterBox sensor/ actuator node.}
        \label{fig:figsensornode}
    \end{subfigure}
    \quad
    \begin{subfigure}[b]{0.49\textwidth}
        \includegraphics[width=\textwidth]{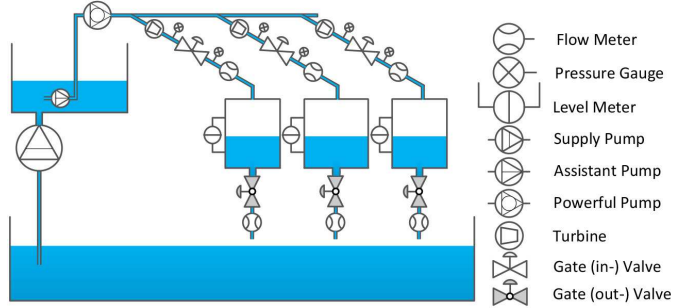}
        \caption{The schematic structure of the WaterBox.}
        \label{fig:waterboxstructure}
    \end{subfigure}
    \caption{WaterBox evaluation platform.}
    \label{fig:waterboxplatform}
\end{figure*}

Smart water networks have been used as a proof of concept for our proposed framework. The WaterBox platform (see Figure \ref{fig:waterboxtestbed}) is a scaled version of such water network \cite{kartakis2015waterbox} and developed to demonstrate real time monitoring and control by adapting innovative communication theories and control methodologies. WaterBox was %extended compared with an early version \cite{kartakis2015waterbox}, in order to be
used as evaluation platform for our proposed ETC techniques and communication schemes. %This section describes the main changes which involve synchronization mechanism, pumping infrastructure, energy harvesting and consumption monitoring system.

\subsection{WaterBox infrastructure}
A Water supply network structure consists of three individual layers: (a) storage and pumping, (b) water supply zones and District Meter Areas (DMAs), and (c) end users (water demand). While valves control flows and pressures at fixed points in the water network, pumps pressurise water to overcome gravity and frictional losses along supply zones, which are divided into smaller fixed network topologies (in average 1500 customer connections) with permanent boundaries, DMAs. The DMAs are continuously monitored with the aim to enable proactive leakage management, simplistic pressure management, and efficient network maintenance. WaterBox was designed to support this architecture as follows:
% \begin {figure}[!ht]
% \centering
% \includegraphics[width=0.9\linewidth]{./Figures/waterboxstructure3}
% \caption{The schematic structure of the WaterBox.}
% \label{fig:waterboxstructure}
% \end {figure}
\subsubsection{Water Storage and Pumping} The structure of the WaterBox is shown as Fig. \ref{fig:waterboxstructure}. The WaterBox has a lower tank (i.e. ground, soil), an upper tank (i.e. reservoir, lake) and three small tanks (i.e. DMAs). The lower tank collects water from small tanks, and supplies water to the upper tank by an underwater bilge supply pump. This supply pump can supply enough water as the system requires. An assistant bilge pump and a new powerful pump are installed in series inside and after the upper tank respectively, and supply water to the small tanks. When the small tanks require more water supply, the assistant pump and powerful pump work together. When the small tanks require less water supply, only the assistant weak pump works. This behaviour emulates the day and night pumping operation of a water network in which the demand changes dramatically. %The pumps are controlled by a separate micro development board which maintains an HTTP server and enables the pump on/off actuation over the network.
% \begin {figure}[!ht]
% \centering
% \includegraphics[width=0.7\linewidth]{./Figures/sensornode.eps}
% \caption{WaterBox sensor/ actuator node.}
% \label{fig:figsensornode}
% \end {figure}
\subsubsection{Water Supply \& Sensor/Actuator Node} The water from the powerful pump flows into three small 'DMA' tanks via a pipe network. For the inlet each tank, a sensor/actuator node (see Figure \ref{fig:figsensornode}), based on the Intel Edison development board \cite{inteledison2015}, controls the water flow though a motorized gate valve, so-called in-valve, and monitors the water flow, pressure (before and after in-valve) and the in-tank water level. Further, a turbine (flow-based energy harvester) is installed before each in-valve to harvest energy. To capture the total energy consumption of each sensor/actuator node, a custom made sensor module was created. % which consists of a separate MCU.
%In spite the capability of battery connection, each sensor/actuator node is connected constantly to power to ensure the robustness of experiments.

\begin{remark} In the WaterBox infrastructure, the sensors and actuator of each inlet are connected to the same node. However, our proposed communication schemes can be applied to non-collocated infrastructures.
\end{remark}

%The inlet of each tank can be monitored and controlled by the sensor/actuator node (see Figure \ref{fig:figsensornode}) which is based on the Intel Edison development board \cite{inteledison2015}. The Intel Edison board incorporates Bluetooth and WiFi communication modules while hosts the Ubilinux operating system \cite{ub2016linux}. Under this software environment, a C++ application was developed to retrieve and log data from the sensors, control the actuators, and perform the communication based on our different ETC scenarios and TDMA-schemes. The flow to each small tank is controlled by a motorized gate valve. These valves are called in-valves. Before and after each in-valve, two new industrial heavy duty pressure sensors \cite{honeywell2015} were installed. The sensors are introduced to monitor the pipe pressure and protect the pipe from burst or leakages. A turbine (flow-based energy harvester) is installed before each in-valve to harvest energy; a flow meter is installed after each in-valve. These meters measure the total volume of the water that has passed the pipes. In the roof of each small tank, a distance sensor was installed. These sensors, which can act as level meters, monitor the heights of the water levels in each small tank. To capture the energy consumption of each sensor/actuator node, a custom made high frequency sensor module was created which consists of a separate MCU. In spite the capability of battery connection, each sensor/actuator node is connected constantly to power to ensure the robustness of experiments.

\subsubsection{Water Demand Emulation} At the bottom of each small tank, there is an opening which enables the emulation of water consumption. A gate valve, so-called out-valve, is installed after each opening and can be controlled by a sensor/actuator node. The control of out-valves changes the outlet flow rate and facilitates the emulation of user's random water consumptions (i.e. external disturbance).

\begin {figure}[!ht]
\centering
\includegraphics[width=0.8\linewidth]{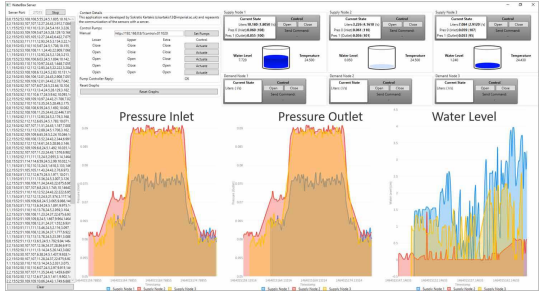}
\caption{Local controller visualization application.}
\label{fig:figlocalcontroller}
\end {figure}

\subsubsection{Base Station (Controller)} Every sensor node is connected to a local isolated WiFi network \footnote{The isolation was achieved by disabling SSID operation (broadcast of WiFi availability to new users) from the router and selecting the low occupied communication channel for the nodes based on spectrometer experiments}. A laptop is used as a controller or base station and hosts a visualization application (see Figure \ref{fig:figlocalcontroller}) which presents at real time the current state of the system, allows the manual control of actuators, logs the retrieved messages, and enables our proposed communication schemes and ETC scenarios per experiment. Additionally, due to lack of an indoor GPS time synchronization, a NTP server is running in the local controller and ensures less than millisecond time synchronization accuracy among the nodes. %Each node activates a NTP client demon as a background task which acts only during the idle or sleeping mode. The active demons and the background task of the operating system cause high energy consumption and for this reason each sensor node logs the sleep time and the corresponding energy consumption separately for further analysis.

\subsection{System Identification \& Modelling}

We apply Grey-Box identification \cite{ljung1999system} to generate the system model and to find the uncertain parameters. A first principles model is obtained following \cite{walski2003advanced}. We identify independently models under both mode 1: only the assistant pump works and mode 2: both pumps work. Using the Matlab cftool, we generate fitting curves for the gate valve coefficient of each in-valve, the turbine efficiency, and the pump efficiency. These curves are used to compute the first principles model. Since our aim is to stabilize the water level of each tank $j\in\{1,2,3\}$ at the desired height $h_j'$, the model is linearised around this height. %the reference water level $h_i'$, which will be discussed later in Section \ref{section:controller}.
%Based on observation, for the given working water levels and constant $\alpha_i^{out}$, defined as open level of each out-valve, the out flow rates remain almost constant, because of the special structure of the WaterBox.
%To simplify the simulation of the user water consumption, we keep $\alpha_i^{out}=360^{\circ}$, thus, constant out flow rates can be assumed.
In this process, in order to simplify the simulation of the user water consumption, we keep the out-valves open, thus, constant out flow rates can be assumed.

\section{Hybrid Controller Design}
%\section{Hybrid Controller \& Triggering Mechanism Design}
\label{section:controller}

%In this section, we design a hybrid controller based on a realistic water network scenario applied on the WaterBox.

%\subsection{Control Scenario}
%The WaterBox infrastructure provides a flexible evaluation platform for realistic water network scenarios.
To evaluate the proposed ETC-based communication schemes, the following %water network control scenario was used: \textit{"Control water network valves and pumps with such a way to provide the appropriate amount of water to each DMA by: (a) covering QoS requirements (pressure and flow lower bounds), (b) ensuring robust water network operation without leakages and bursts (pressure and flow upper bounds), and (c) minimizing pumping energy consumption though an efficient scheduling by exploiting demand changes along the time."}
%To apply this scenario to WaterBox, the above can be translated as follows:
\textbf{control scenario} was used: \textit{"Control in-valves to stabilize the water level of the small 'DMA' tanks to a certain level ensuring pressure and flow bounds. Enable mode 2 (weak and powerful pump) only if the system is away from the target levels. Switch to mode 1 (weak pump) only when the system is close to the reference state to guarantee efficient low pressure and flow operation."} %By having the control scenario definition, the section continues with the formal design of a hybrid controller.

\subsection{Hybrid Controller Design}
In the design of the controller, the following limitations need to be considered:
\begin{enumerate}
  \item \textit{Saturation of the actuators:} The maximum open level of one in-valve is $360^{\circ}$, while the minimum is $0^{\circ}$.
  \item \textit{Actuator quantization:} Due to the limitation of the valve's control components, their open levels can only be changed in steps of $10^{\circ}$. Therefore, small disturbances may result in dramatic changes of actuations.
  \item \textit{Over-pressure protection:} Due to mechanical limitations of the pipe network, there is a maximum allowable pressure for the pipe network.
\end{enumerate}

Since the height of the water levels have a direct effect on the Quality of Service (QoS), the closed-loop system requires a fast response; however, since the size of the small tanks are limited, the overshoot should simultaneously be constrained. Experiments show that, in mode 2, the pipe network may be over pressured, when the open level of the in-valves, defined by $\alpha_j^{in}$, cannot satisfy:
\begin{equation}\label{eq:minimumopenlevel}
\sum_{j=1}^3\alpha_j^{in}\geq 180^{\circ}.
\end{equation}
Overshoot and disturbances could make condition (\ref{eq:minimumopenlevel}) be violated. While in mode 1, there is no such constrains, that is, even all three in-valves are totally closed, the pipe network will not be over pressured. Therefore, filling the small tanks in mode 2 and switching to mode 1 when (\ref{eq:minimumopenlevel}) is violated is required. Also experimentally, we observe that, when the system is in mode 1, the pump may not provide enough water supply to the small tanks even at the maximum open level, i.e. $\alpha_j^{in}=360^{\circ},~\forall j\in\{1,2,3\}$. Therefore, switching back to mode 2 when the water in the tanks reaches some pre-defined low levels is necessary. To support this mode switching, we define $\underline{h}:=\begin{bmatrix}
\underline{h}_1 & \underline{h}_2 & \underline{h}_3\\
\end{bmatrix}^{\mathrm{T}},\,\underline{h}_j<h'_j,\,j\in\{1,2,3\}$, as the lower water levels. If $\exists j\in\{1,2,3\}$ such that $h_j(t)\leq\underline{h}_j$, the system switches from mode 1 to mode 2. With carefully chosen $\underline{h}_j$ and properly designed controllers, this violation can only happen in mode 1. Further analysis shows that, (\ref{eq:minimumopenlevel}) can only be violated when $h_j(t)>h_j'$, which together with the fact that $\underline{h}_j<h'_j$ precludes Zeno behaviour.

Let $\vartheta\in\{1,2\}$ represents the corresponding system mode. The  linearised switched model and switched controller of WaterBox are described by:
\begin{equation}\label{eq:waterboxmodel}
\dot\xi(t)=A\xi(t)+B_\vartheta v(t),\,\vartheta=1,2,
\end{equation}
\begin{equation}\label{eq:switchedcontroller}
v(t)=S(-K_\vartheta\xi(t)+\bar\alpha_\vartheta^{in}),\,\vartheta=1,2,
\end{equation}
%\begin{equation}\label{eq:switchedcontroller}
%v(t)=\left\{
%\begin{aligned}
%&-K_1\xi(t)+\bar\alpha_1^{in}, \text{ when }j=1\\
%&-K_2\xi(t)+\bar\alpha_2^{in}, \text{ when }j=2,\\
%\end{aligned}\right.
%\end{equation}
% \begin{equation}\label{eq:switchlaw}
% \left\{
% \begin{aligned}
% &j=1\to j=2 \text{ if } \lor_{i=1}^3 \xi_i(t)\leq \underline{h}_i-h'_i\\
% %&j=2\to j=1 \text{ if } \sum_{i=1}^3v_i< 180^{\circ},
% &j=2\to j=1 \text{ if } |S(-K_2\xi(t)+\bar\alpha_2^{in})|_{1}<180^{\circ},
% \end{aligned}\right.
% \end{equation}

where $\xi(t)=\begin{bmatrix}
\xi_1(t) & \xi_2(t) & \xi_3(t) \\
\end{bmatrix}^{\mathrm{T}}$, $\xi_j(t):=h_j(t)-h_j',\,j=1,2,3$ are the system states, $h_j(t)\in\mathbb{R}$ are the water levels, and $h_j'\in\mathbb{R}$ are the reference water levels; $v(t)=\begin{bmatrix}
v_1(t) & v_2(t) & v_3(t) \\
\end{bmatrix}^{\mathrm{T}}$, $v_j(t):=\alpha_j^{in},\,j=1,2,3$ are the system control inputs, %with the open levels $\alpha_i^{in}=[0, 360]$; %the open levels of the in-valve
and $\bar\alpha_\vartheta^{in}$ are the equilibrium open levels of the in-valves per operation mode $\vartheta$; $S$ is a map $\mathbb{R}^{m}\rightarrow\mathbb{R}^{m}$ representing actuator saturation and quantization, that is $S(s_j):=\max\{\min\{10\lfloor 10^{-1}s_j\rfloor,360^{\circ}\},0^{\circ}\}$. %$v_i(t):=\alpha_i^{in},\,i=1,2,3$ are the system control inputs, %with the open levels $\alpha_i^{in}=[0, 360]$; %the open levels of the in-valve and $\bar\alpha_j^{in},\,j=1,2$ are the equilibrium open levels of the in-valves.

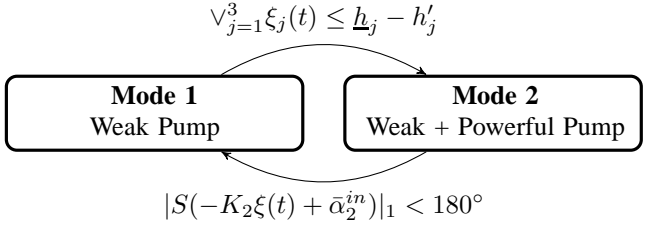
\begin{figure}[!ht]
\begin{tikzpicture}[->,>=stealth']

\node[state, text width=3.8cm] (QUERY)
 {\begin{tabular}{c}
  \textbf{Mode 1}\\
  Weak Pump
  \end{tabular}};

 \node[state,    	% layout (defined above)
  text width=3.8cm, 	% max text width
  right of=QUERY, 	% Position is to the right of QUERY
  node distance=4.5cm, 	% distance to QUERY
  anchor=center] (ACK) 	% posistion relative to the center of the 'box'
 {%
 \begin{tabular}{c} 	% content
  \textbf{Mode 2}\\
  Weak + Powerful Pump
 \end{tabular}
 };

 \path (QUERY) edge[bend left]  node[anchor=south,above]{$\lor_{j=1}^3 \xi_j(t)\leq \underline{h}_j-h'_j$} (ACK)

(ACK) edge[bend left]  node[anchor=north,below]{$|S(-K_2\xi(t)+\bar\alpha_2^{in})|_{1}<180^{\circ}$} (QUERY);

\end{tikzpicture}
\caption{Hybrid controller state automaton.}
\label{fig:figstateautomaton}
\end {figure}

Then, the WaterBox hybrid controller state automaton is illustrated in Figure~\ref{fig:figstateautomaton}.

% \begin {figure}[!ht]
% \centering
% \includegraphics[width=0.7\linewidth]{./Figures/switchSM.eps}
% \caption{Hybrid controller state automaton.}
% \label{fig:figstateautomaton}
% \end {figure}

%\bigwedge_{i=1}^3h_j\geq \underline{h}_j
%\sum_{i=1}^3\alpha_i^{in}\geq 180^{\circ}
From Grey-Box identification procedure, the system parameters are: $A$ is a zero $3\times 3$ matrix and
\begin{equation*}
\begin{aligned}
% A=\scriptsize{\begin{bmatrix}
% 0 & 0 & 0 \\
% 0 & 0 & 0 \\
% 0 & 0 & 0 \\
% \end{bmatrix}},
B_1=10^{-5}\times\scriptsize{\begin{bmatrix}
    0.1436  & -0.0170 &  -0.0164 \\
   -0.0098  &  0.1060 &  -0.0100 \\
   -0.0139  & -0.0139 &   0.1492 \\
\end{bmatrix}},\\
~~B_2=10^{-5}\times\scriptsize{\begin{bmatrix}
    0.7666  & -0.0493 &  -0.0457 \\
   -0.0274  &  0.5848 &  -0.0279 \\
   -0.0393  & -0.0432 &   0.7865 \\
\end{bmatrix}}.
\end{aligned}
\end{equation*}
The controllers designed are given by:
\begin{equation*}
\begin{aligned}
K_1=\scriptsize{\begin{bmatrix}
99950  &  3029  &  872   \\
-3014  &  99940 &  -1679 \\
-922   &  1652  &  99982 \\
\end{bmatrix}},
K_2=\scriptsize{\begin{bmatrix}
9998.5  &  167.1  &  41.0   \\
-166.6  &  9997.9 &  -116.0 \\
-43.0   &  115.3  &  9999.2 \\
\end{bmatrix}}.
\end{aligned}
\end{equation*}

The designed controller is stable in both mode 1 and 2 because $-B_1K_1$ and $-B_2K_2$ are Hurwitz matrices. Further,
%Therefore, with the designed controller, each of the mode are stable.
due to the long dwell time of the system, the closed loop retains stability. Given $h_j'=0.06$ and $\underline{h}_j=0.03$, $\forall j\in\{1,2,3\}$, $\bar\alpha_1^{in}$ and $\bar\alpha_2^{in}$ are computed:
\begin{equation*}
\bar\alpha_1^{in}=\scriptsize{\begin{bmatrix}
503.5950 \\
422.4378 \\
428.5839 \\
\end{bmatrix}},\,\bar\alpha_2^{in}=\scriptsize{\begin{bmatrix}
84.5099 \\
68.2069 \\
72.8442 \\
\end{bmatrix}}.
\end{equation*}
%One can easily find that, around the reference point, since the maximum open level is $360^{\circ}$, the actual system cannot converge to the reference point in mode 1, as we have pointed out. In mode 2, the sum of each element in $\bar\alpha_2$ is $225.561^{\circ}$, which is bigger than $180^{\circ}$. This make it possible for mode 2 to converge to the reference point. However, because of the overshoot and actuator quantization, the system may probably switch to mode 1 during converging.

\section{Evaluation}
\label{section:Evaluation}
This section summarizes % aggregates
the experimental results of more than 400 experiments conducted in WaterBox to evaluate our proposed communication schemes for the different ETC strategies. Each experiment executes the same control scenario (as described in Section \ref{section:controller}) and the total process lasts between 7 and 10 minutes, including the water state initialization, the execution of experiment, and data logging from sensor/ actuator nodes and local controller. %Before the data analysis, we provide the evaluation setup of both TDMA schemes and ETC approaches.

\subsection{Evaluation Setup}

\begin{table}[ht]
\caption{Communication parameter evaluation setup.}
\label{table:comparametersetup}
\centering
\resizebox{\linewidth}{!}{\begin{tabular}{c|c|c|l|}
\cline{2-4}
\multicolumn{1}{l|}{} & \textbf{Parameter} & \textbf{Value} & \textbf{Description} \\ \hline \hline
\multicolumn{1}{|c|}{\multirow{4}{*}{\textbf{\begin{tabular}[c]{@{}c@{}}Packet\\ Size\end{tabular}}}} & $x_j$ & 36 Bytes & \begin{tabular}[c]{@{}l@{}}Node ID\\ Timestamp (in msec)\\ Inlet pressure\\ Outlet pressure\\ Flow rate\\ Total water volume\\ Distance from surface\\ Energy consumption\\ Energy harvesting\end{tabular} \\ \cline{2-4}
\multicolumn{1}{|c|}{} & $Ack$   |   $r_j$, $a_j$ & 1 Byte & 0 or 1 |  Node ID \\ \cline{2-4}
\multicolumn{1}{|c|}{} & $u_j$, $\eta _j$, $\theta _j$ & 2 Bytes & \begin{tabular}[c]{@{}l@{}}Control signal and \\ DETC parameters\end{tabular} \\ \cline{2-4}
\multicolumn{1}{|c|}{} & $m_j$ & 4 Bytes & State delta \\ \hline \hline
\multicolumn{1}{|c|}{\multirow{6}{*}{\textbf{\begin{tabular}[c]{@{}c@{}}Time\\ Duration\end{tabular}}}} & $S_j^x$ & 80 msec & X slot size \\ \cline{2-4}
\multicolumn{1}{|c|}{} & $S_j^u$, $S_j^v$ & 50 msec & U and V slot size \\ \cline{2-4}
\multicolumn{1}{|c|}{} & $d_c$ & 10 msec & Control decision delay \\ \cline{2-4}
\multicolumn{1}{|c|}{} & $d_g$ & 5 msec & \begin{tabular}[c]{@{}l@{}}Threshold violation \\ decision delay\end{tabular} \\ \cline{2-4}
\multicolumn{1}{|c|}{} & Guard slot & 1 msec & \begin{tabular}[c]{@{}l@{}}Forced task \\ termination time\end{tabular} \\ \hline
\end{tabular}}
\end{table}

%In Section \ref{section:MAC}, we proposed the three communication protocols, C-TDMA, ADC-TDMA, and SDC-TDMA that accommodate TTC and PETC, PADETC, and PSDETC scenarios respectively.
The proposed communication protocols of Section \ref{section:MAC} were deployed to the WaterBox sensor/actuator nodes by wrapping the functionality of the Intel Edison WiFi module. Table \ref{table:comparametersetup} presents the configuration of the communication parameters. Based on the predefined packet sizes of the specific hardware infrastructure, a set of experiments was conducted to determine reliable time slot and guard delay sizes.

\begin{table}[ht]
\caption{Parameters of triggering strategies%: TTC (time-triggered), PETC (periodic), PSDETC (periodic synchronous decentralized), PADETC (periodic asynchronous decentralized).
.}
\label{table:parametersofdifferentstrategies}
\centering
{\begin{tabular}{|@{\hspace{12pt}}c@{\hspace{12pt}}|@{\hspace{20pt}}c@{\hspace{20pt}}|c|@{\hspace{15pt}}c@{\hspace{15pt}}|}
\hline
\textbf{Method} & $\mathbf{T}$ \textbf{(sec)} & \textbf{Parameter} & \textbf{Value} \\ \hline \hline
TTC & \textbf{0.5}, 1, \textbf{2} & - & - \\ \hline
PETC & \textbf{0.5}, 1, \textbf{2} & \multirow{2}{*}{$\sigma$} & \multirow{2}{*}{0.05, 0.1, \textbf{0.2}} \\ \cline{1-2}
PSDETC & 1, \textbf{2} &  &  \\ \hline
\multirow{2}{*}{\begin{tabular}[c]{@{}c@{}}PADETC\\ (abs \& rel)\end{tabular}} & \multirow{2}{*}{\textbf{0.5}, 1, \textbf{2}} & $\mu$ & 0.75, \textbf{0.95} \\ \cline{3-4}
 &  & $\varrho$ & \textbf{85}, 120 \\ \hline
\end{tabular}}
\end{table}

Based on the Table \ref{table:comparametersetup} timing parameters and the Section \ref{section:MAC} time slot analysis, the minimum interval size $T_{min}$ for C-TDMA and and ADC-TDMA has to be more than 321 msec while for SDC-TDMA more than 564 msec (because of the V-slots). %By having these assumptions and in order to ensure reliable operation, we decided to
Thus, we evaluate TTC, PETC, and PADETC (with absolute or reference value) with interval size $T=0.5, 1, 2$ %while for
and $T=1, 2$ sec for PSDETC. The selected interval sizes and the rest parameters of the ETC strategies are listed in Table \ref{table:parametersofdifferentstrategies}. The $\sigma$ and $\rho$ ETC parameters are chosen by finding feasible solutions of the corresponding algorithms (\ref{eq:petclmi}) and (\ref{eq:paetclmi}), while $\mu$ is tuned experimentally.
%The ETC parameters are chosen either by finding feasible solutions to the corresponding algorithm: $\sigma$ by (\ref{eq:petclmi}), $\rho$ by (\ref{eq:paetclmi}), or by carefully chosen, such as $\mu$.

In the first set of experiments, we examine the impact of the ETC parameters ($\sigma$, $\mu$, and $\varrho$) to the performance of the system. A fixed interval size $T=1$ was used with all the different combinations of ETC parameter values of Table \ref{table:parametersofdifferentstrategies}. Another set of experiments was conducted to explore the effect of $T$ in the behaviour of the system. Keeping $\sigma=0.2$, $\mu=0.95$, and $\varrho=85$ constant, all the experiments were re-executed with $T=0.5$ and $T=2$ (Table \ref{table:parametersofdifferentstrategies} bold text). To ensure the validity of the evaluation results, each experiment was repeated 10 times\footnote{The number of experiment repetitions was selected experimentally by analysing the variance of the results (i.e. under 2\% of mean).} for each different combination of ETC parameters and $T$. Mean values are used to illustrate the evaluation results. The data was captured from the nodes and controller for the period of time between the beginning of each experiment ($t_0=0$) until a fixed end time ($t_{end}=110s$), which guarantees the system turns to mode 1 and converges to steady state, denoted $T_{exp}$.

\subsection{Experimental Results}
\begin {figure}[!ht]
\centering
\includegraphics[width=1\linewidth]{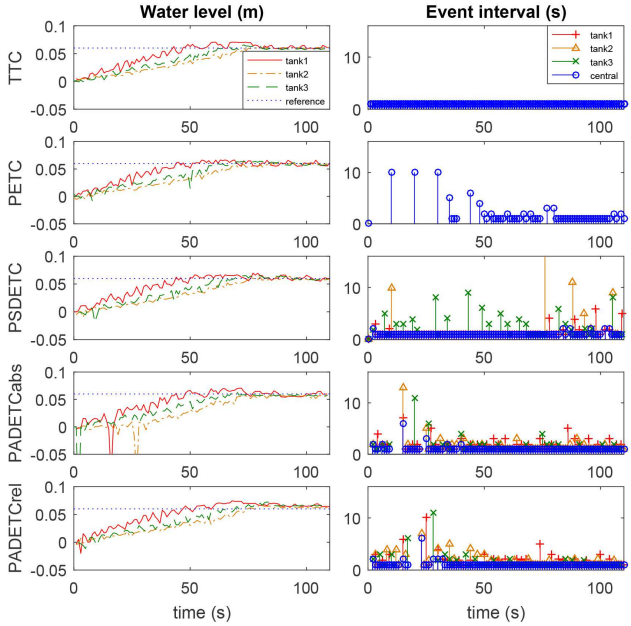}
\caption{Experimental results. Note that, for SETC, the maximum event-interval of small tank 2 is 66 sec.}
\label{fig:experimentresults}
\end {figure}

In this section we compare TTC, PETC, PSDET, PADETabs and PADETrel experimental results, in terms of:

\begin{figure*}[!ht]
	\centering
    \begin{subfigure}[b]{0.24\textwidth}
        \includegraphics[width=\textwidth]{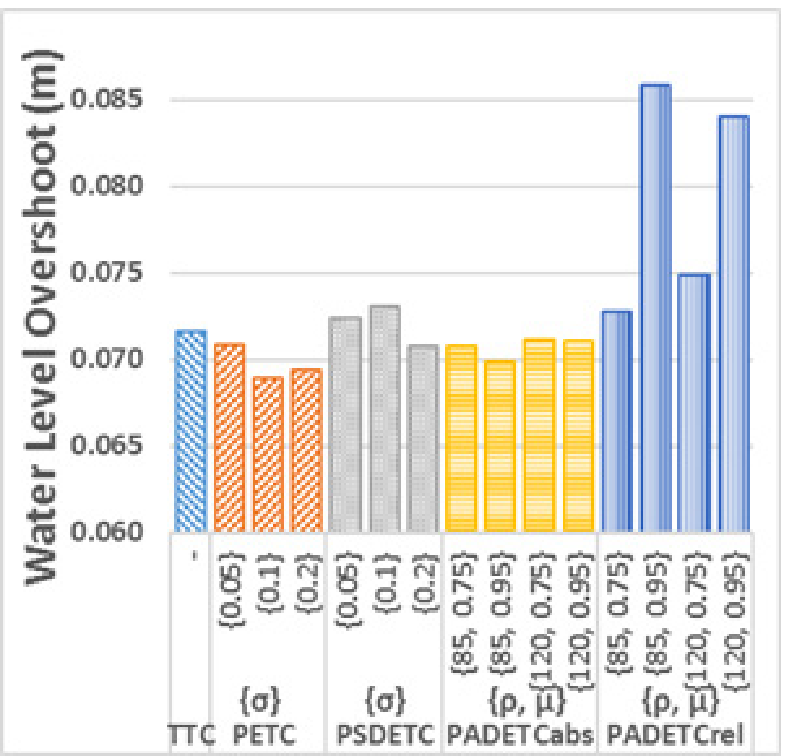}
        \caption{}
        \label{fig:figEwle}
    \end{subfigure}
    \begin{subfigure}[b]{0.24\textwidth}
        \includegraphics[width=\textwidth]{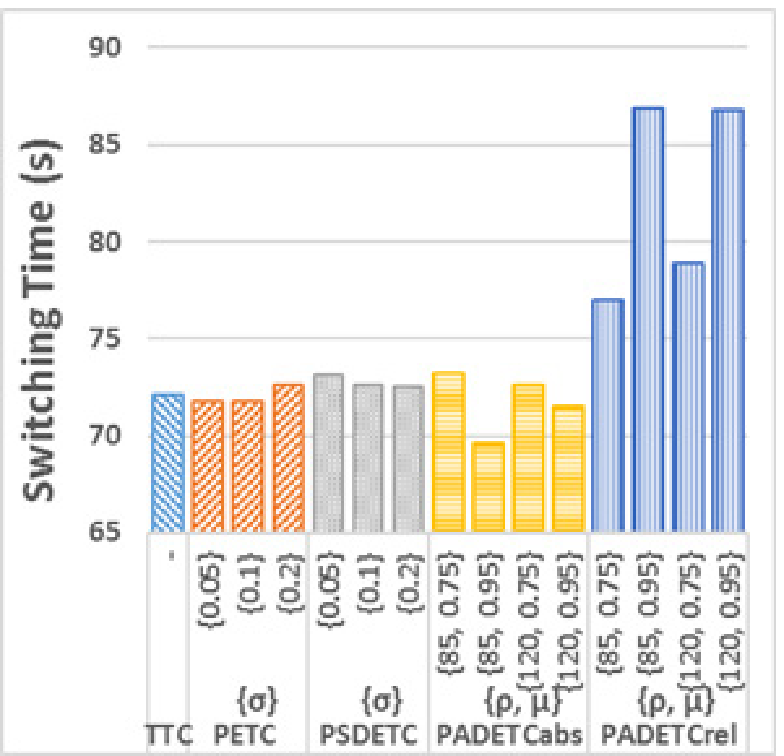}
        \caption{}
        \label{fig:figEct}
    \end{subfigure}
    \begin{subfigure}[b]{0.24\textwidth}
        \includegraphics[width=\textwidth]{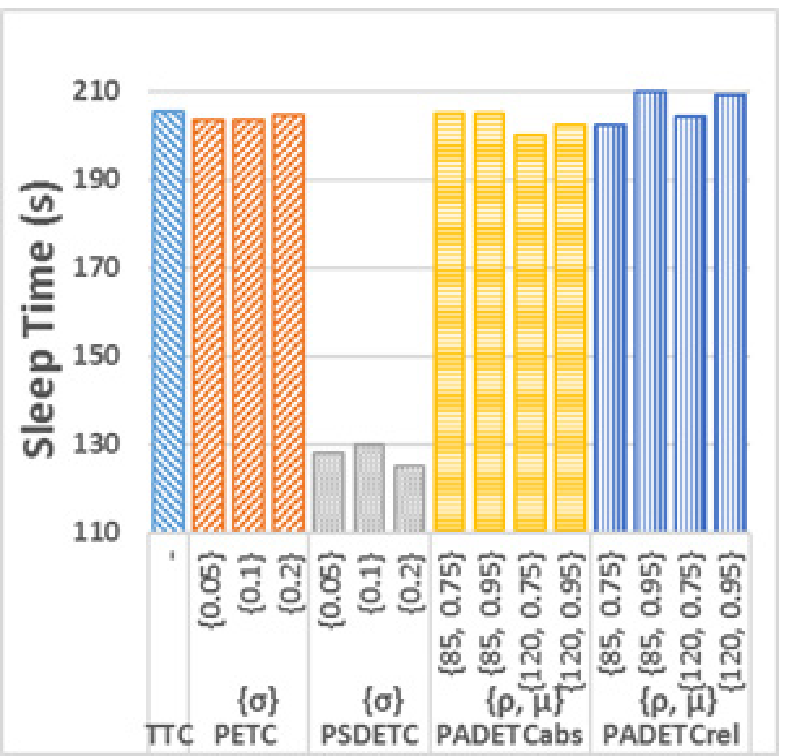}
        \caption{}
        \label{fig:figEst}
    \end{subfigure}
    \begin{subfigure}[b]{0.24\textwidth}
        \includegraphics[width=\textwidth]{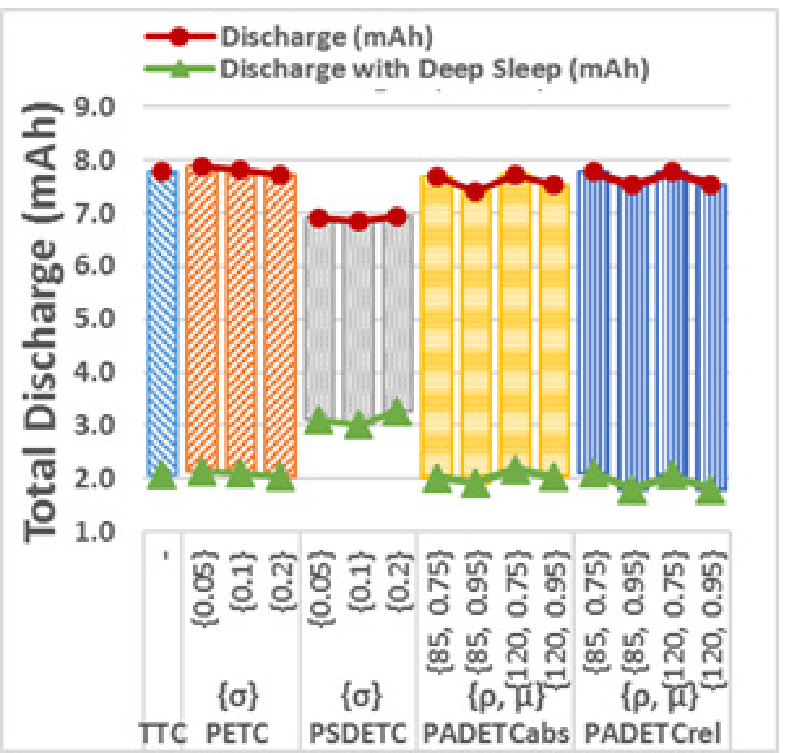}
        \caption{}
        \label{fig:figEd}
    \end{subfigure}
    \begin{subfigure}[b]{0.24\textwidth}
        \includegraphics[width=\textwidth]{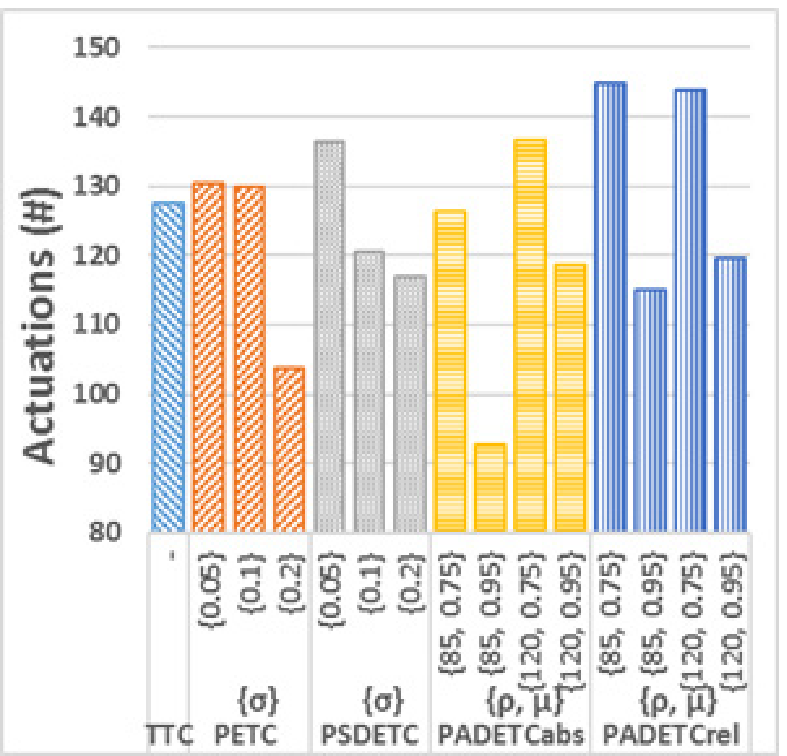}
        \caption{}
        \label{fig:figEact}
    \end{subfigure}
    \begin{subfigure}[b]{0.24\textwidth}
        \includegraphics[width=\textwidth]{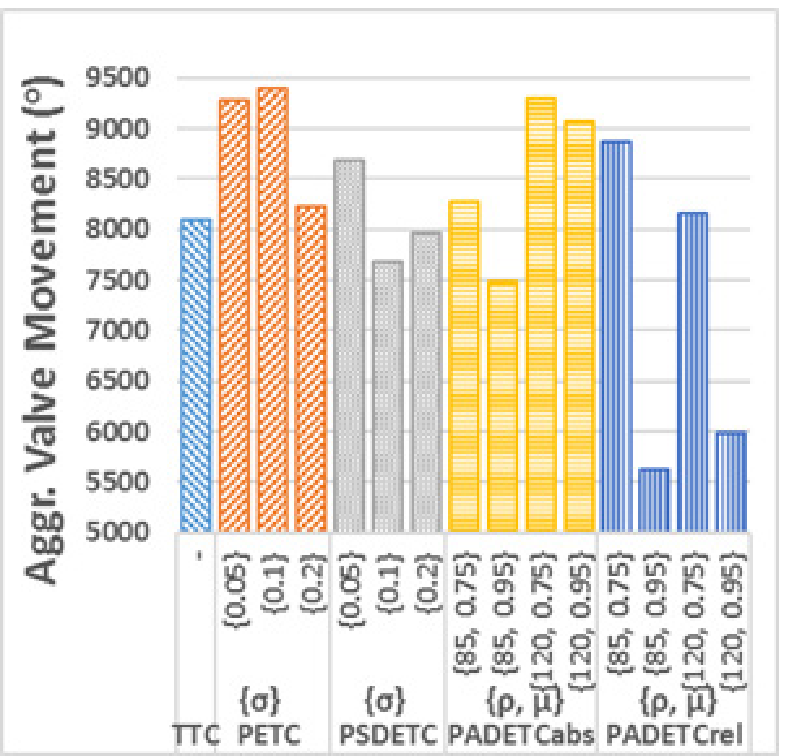}
        \caption{}
        \label{fig:figEavm}
    \end{subfigure}
    \begin{subfigure}[b]{0.24\textwidth}
        \includegraphics[width=\textwidth]{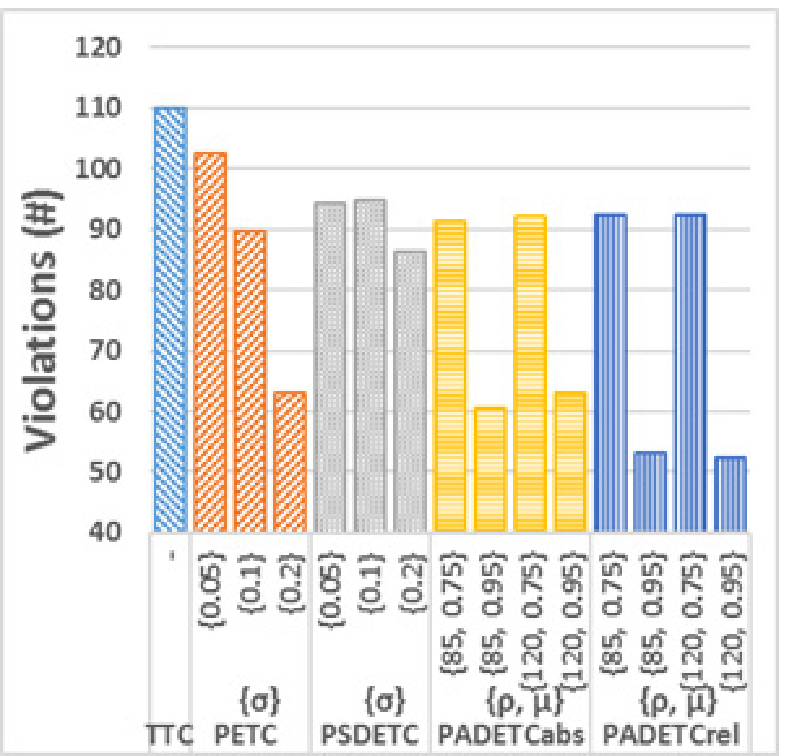}
        \caption{}
        \label{fig:figEv}
    \end{subfigure}
    \begin{subfigure}[b]{0.24\textwidth}
        \includegraphics[width=\textwidth]{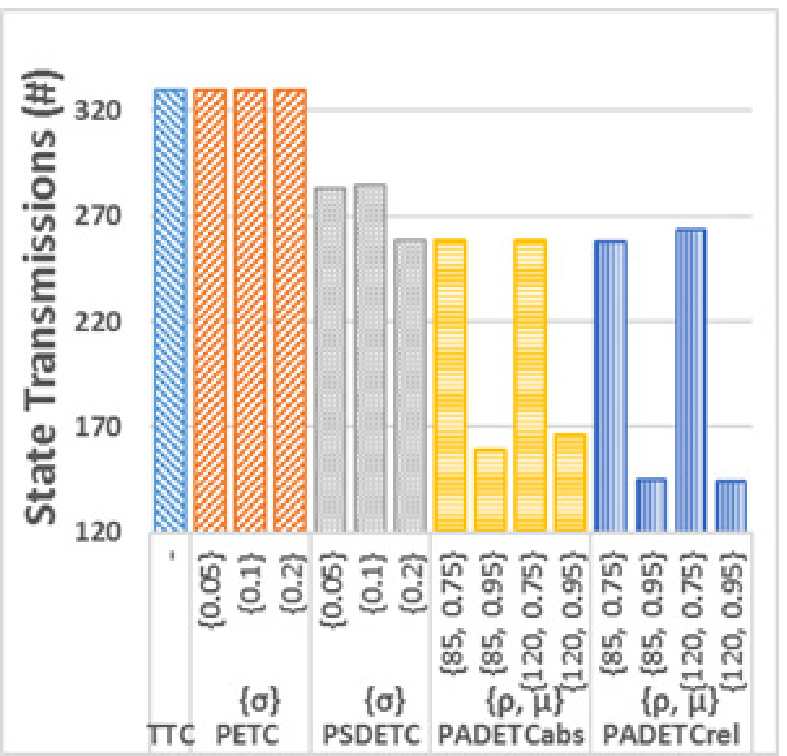}
        \caption{}
        \label{fig:figEtrans}
    \end{subfigure}
    \caption{Impact of ETC parameters ($\sigma, \mu, \rho$) in: (a) water level overshoot, (b) switching time, (c) sleep time, (d) discharge, (e) actuations, (f) valve movement, (g) violation, and (h) state transmissions.}
     \label{fig:figEim}
\end{figure*}

\begin{itemize}
\item \textit{Water Level Overshoot:} the maximum water level during the experiment. This parameter indicates the system's maximum state overshoot which is critical for water network asset safe operation.

\item \textit{Switching Time ($t_{sm}$):} the duration between experiment start time $t_{0}$ and first switch mode time $t_{sm}$, as described in condition (\ref{eq:minimumopenlevel}). The time to mode switching is employed as an estimate of the system settling time (due to its ease of detection in our setup).%Instead of convergence time, the switch mode time is employed as a better indicator of the convergence rate. The reason is that, during mode 1, the unstable performance of the weak pump is unable to stabilize the water level exactly to the reference point, while mode switching is unavoidable to every experiment.

\item \textit{Sleep Time:} the total sleep time of all the nodes. This parameter evaluates the use of the bandwidth and CPU in the sensor/actuator node.

\item \textit{Discharge (Energy Consumption):} our custom made sensor module retrieves current measurements $c_n$ in mA at a fixed frequency of 10kHz. The energy consumption of a specific time period $\Delta t$ in seconds and with average current measurements over this period $\big \langle C\big \rangle_{\Delta t}$ can derived from $E(\Delta t)=\big \langle C\big \rangle_{\Delta t} \cdot \frac{\Delta t}{3600}$. We used a hardware average to ensure the continuity of the current measurements and validated our instrument against a calibrated reference \cite{power2016}. The energy consumption includes the consumption because of the communication, sensing, actuation, and idle mode. We present two discharge values, i.e. the whole discharge and discharge without sleep time. Based on these parameters, the battery lifetime of different hardware infrastructures can be implied.

\item \textit{Actuations:} the number of valves' changes, i.e. $\sum_{\forall{T_i}\in [t_{0}, t_{end}]} \sum_{j=1}^{3}\alpha_j^{in}(T_i)$. The amount of actuations indicates the lifetime of actuators which is vital for industrial deployments.

\item  \textit{Valve Movement:} the sum of valves' movement in degrees between two consecutive changes, i.e. $\sum_{\forall{T_i}\in [t_{0}, t_{end}]} \sum_{j=1}^{3}|\alpha_j^{in}(T_i) - \alpha_j^{in}(T_{i-1})|$. Combined with the number of actuations, the valve movement can be used to estimate physical system lifetime.% maintenance cost estimation by calculating accurately energy consumption and predicting physical system fatigue.

\item  \textit{Violations:} the sum of event condition violations. For each violation the local controller transmits a control signal $u_j$ to each node i.e. three times. Therefore, the total transmitted control signal are equal to 3 times the violations. This metric indicates the communication requirements of actuators. Violations and actuations are different values because  the local controller can produce the same consecutive control signal.

\item  \textit{State Transmissions:} the sum of state $x_j$ transmissions to local controller. This metric indicates the communication requirements of sensors. Both violations and state transmissions represents the total communication requirements of the system.
\end{itemize}

Figure \ref{fig:experimentresults} shows an example of raw data as captured from the nodes and the controller. Next, we analyse the energy consumption trends compared to sleep time for different hardware infrastructures, the effect of ETC parameter setup and interval length $T$ to the performance of the system, and we aggregate the savings of ETC approaches against the vanilla scenario of TTC.

\subsubsection{\textbf{Energy Consumption and Sleep Time}}
\begin{figure*}[!ht]
	\centering
    \begin{subfigure}[b]{0.24\textwidth}
        \includegraphics[width=\textwidth]{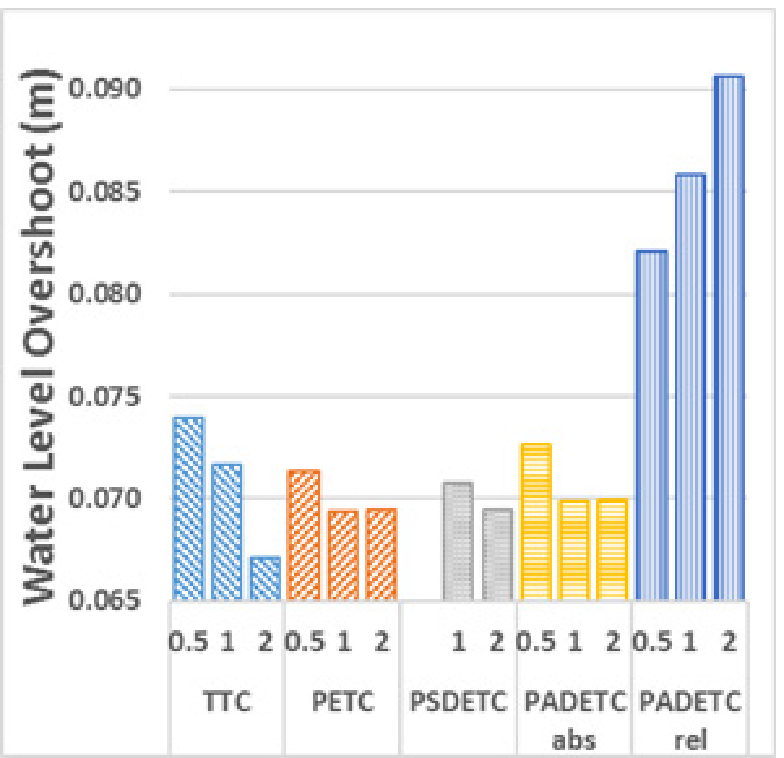}
        \caption{}
        \label{fig:figTwle}
    \end{subfigure}
    \begin{subfigure}[b]{0.24\textwidth}
        \includegraphics[width=\textwidth]{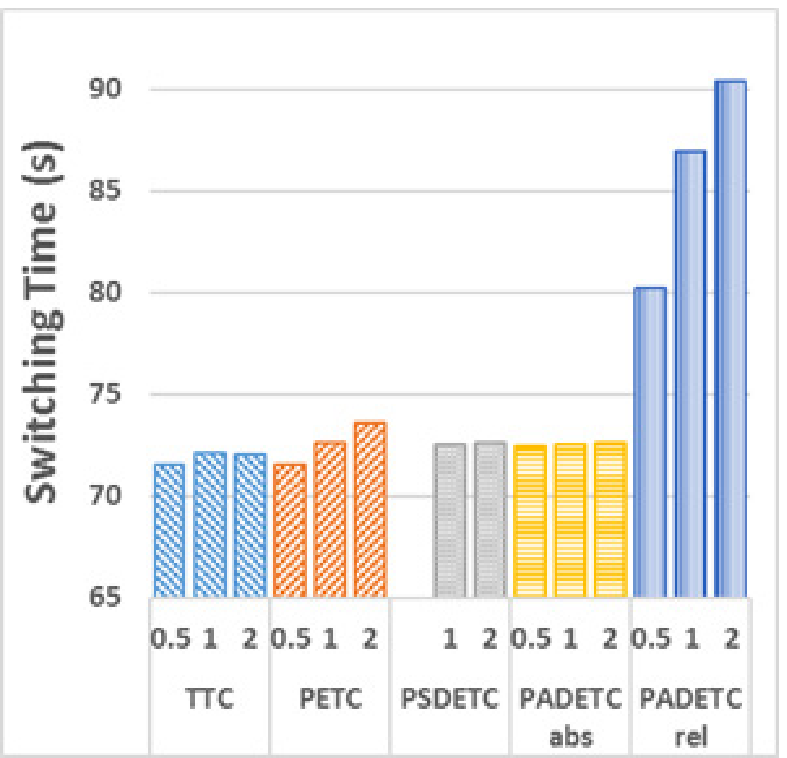}
        \caption{}
        \label{fig:figTc}
    \end{subfigure}
    \begin{subfigure}[b]{0.24\textwidth}
        \includegraphics[width=\textwidth]{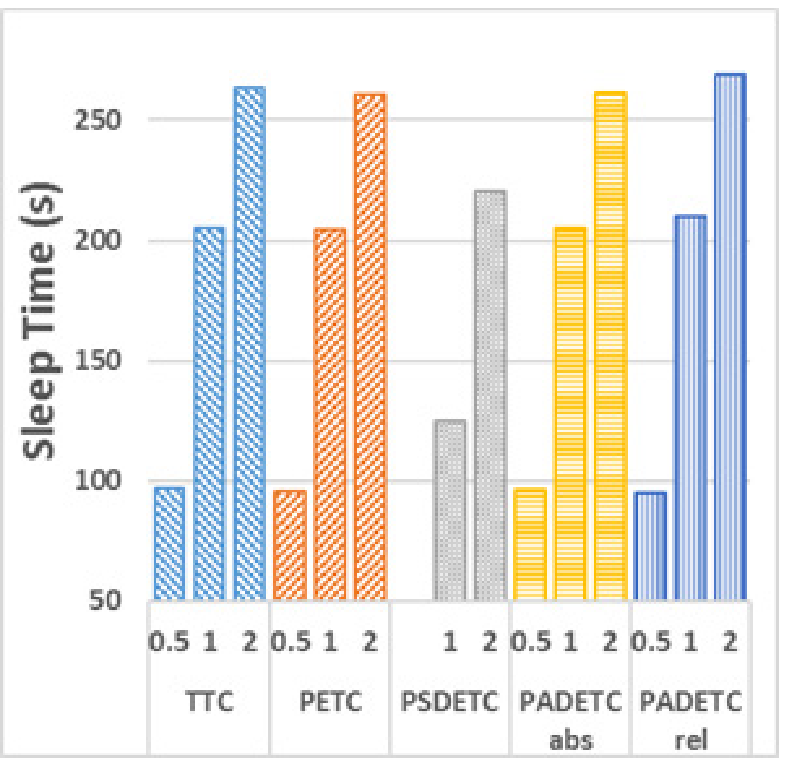}
        \caption{}
        \label{fig:figTst}
    \end{subfigure}
    \begin{subfigure}[b]{0.24\textwidth}
        \includegraphics[width=\textwidth]{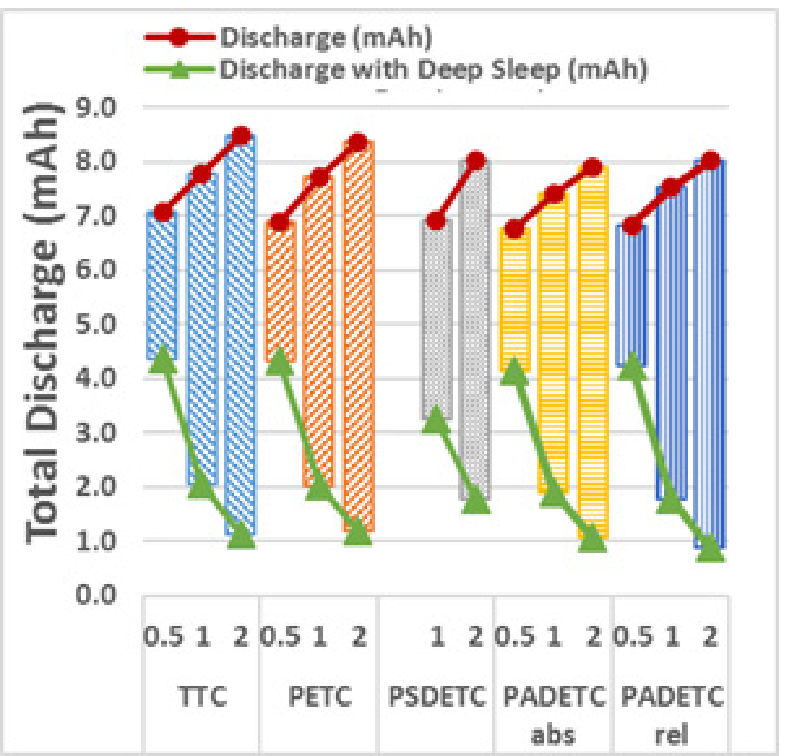}
        \caption{}
        \label{fig:figTd}
    \end{subfigure}
     \begin{subfigure}[b]{0.24\textwidth}
        \includegraphics[width=\textwidth]{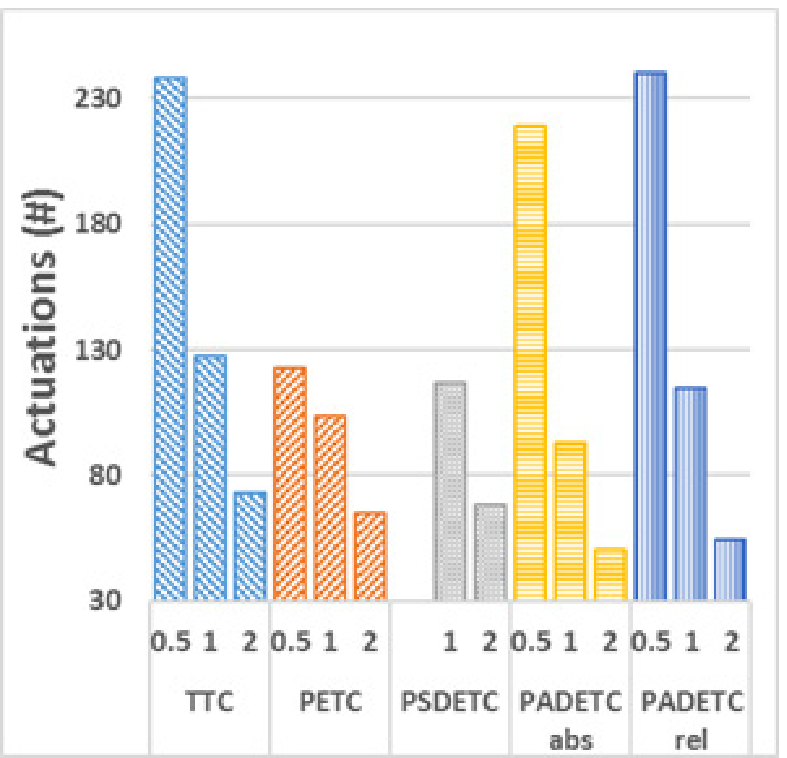}
        \caption{}
        \label{fig:figTact}
    \end{subfigure}
    \begin{subfigure}[b]{0.24\textwidth}
        \includegraphics[width=\textwidth]{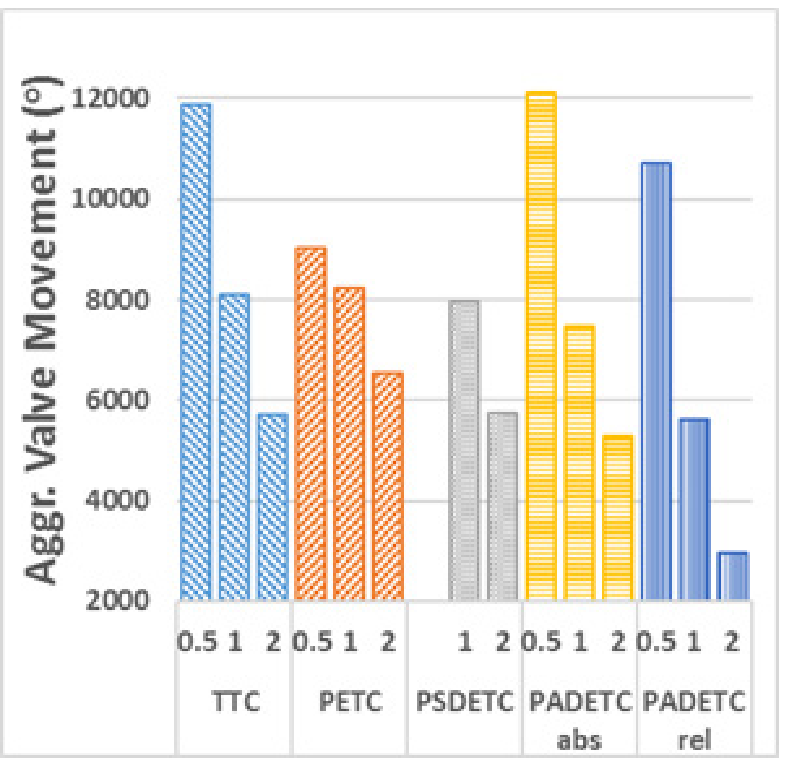}
        \caption{}
        \label{fig:figTavm}
    \end{subfigure}
    \begin{subfigure}[b]{0.24\textwidth}
        \includegraphics[width=\textwidth]{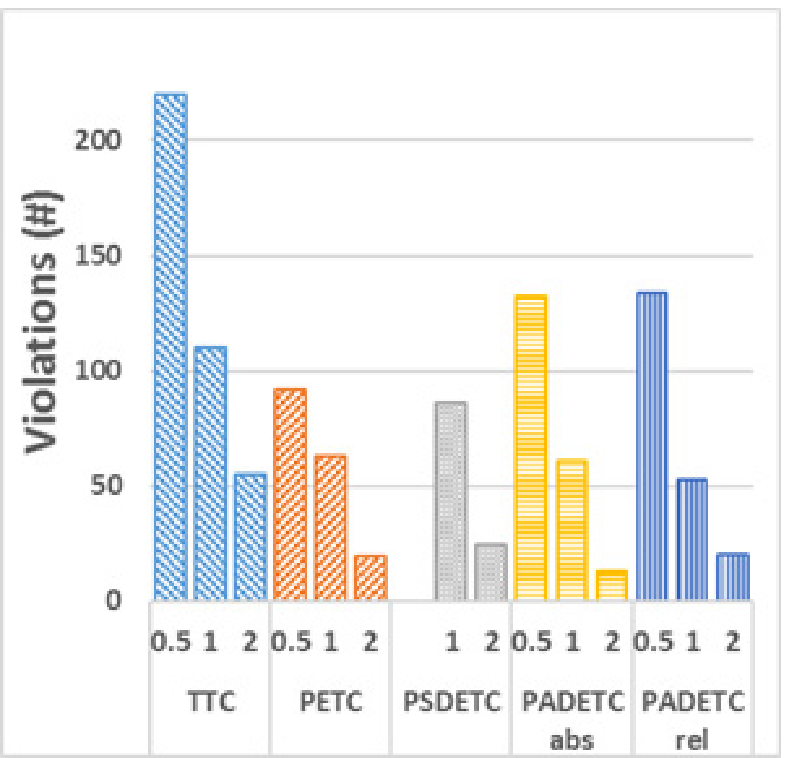}
        \caption{}
        \label{fig:figTv}
    \end{subfigure}
    \begin{subfigure}[b]{0.24\textwidth}
        \includegraphics[width=\textwidth]{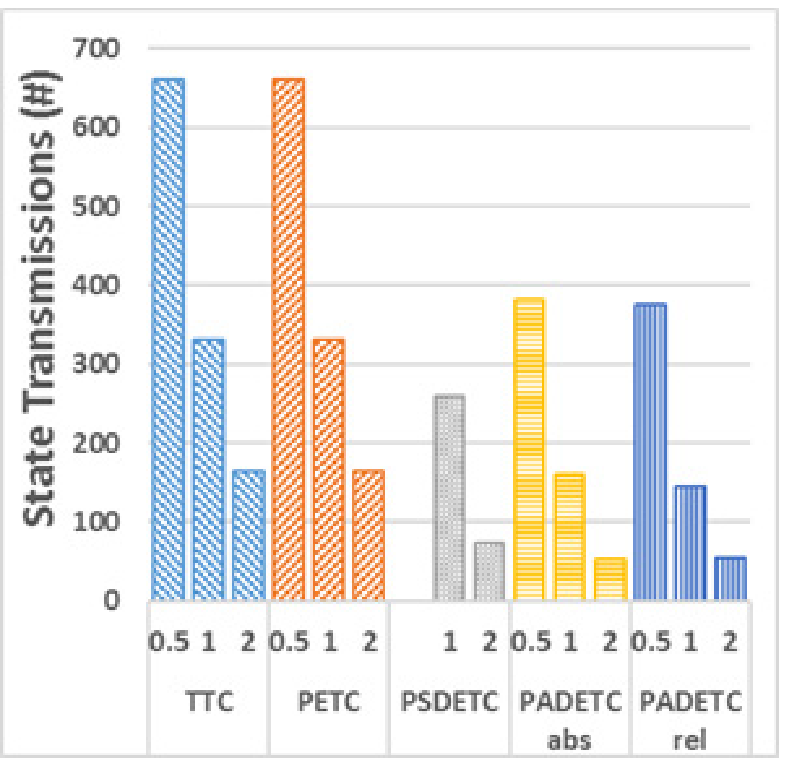}
        \caption{}
        \label{fig:figTtrans}
    \end{subfigure}
    \caption{Impact of interval length $T$ in: (a) water level overshoot, (b) switching time, (c) sleep time, (d) discharge, (e) actuations, (f) valve movement, (g) violation, and (h) state transmissions.}
     \label{fig:figTim}
\end{figure*}

The hardware infrastructure of a WaterBox node consumes more energy in sleep mode. During sleep mode, our process yields priority to the background tasks of the operating system which are more energy hungry. In order to generalize the results to different hardware infrastructures which support lower energy consumption during sleep mode (i.e. deep sleep), we provide the upper and a lower bounds of energy consumption. The upper bound presents the real experimental results based on our node while the lower bound represents an estimation of energy consumption of a node which supports deep sleep\footnote{we calculated the lower energy consumption by subtracting the energy consumption during sleep mode from the total}. The need of energy consumption range can be clearly seen in Figures \ref{fig:figTst} and \ref{fig:figTd}. In spite of the sleep time increase in all cases, the upper bound of energy consumption increases proportionally (the opposite holds for the lower bound). Additionally, PSDETC is expected to consume more energy than the others because of the V-slots. However, Figures \ref{fig:figEst} and \ref{fig:figEd} illustrate the opposite trend for the upper bound (opposite for lower bound) due to the energy hungry sleep mode. Overall, PADETCabs and PADETCrel consume the least energy compared to the other approaches. In spite of the uses of the same communication scheme, PETC performs slightly better than TTC of actuation reduction (quantitative results will presented later on).

%Figures \ref{fig:figEd} and \ref{fig:figTd} illustrate the total energy consumption  of the system, including both communication and actuation, for different ETC parameter setup and interval lengths respectively. In both figures, an upper and lower bound of energy consumption has been defined to indicate the behaviour of different hardware infrastructures for different ETC approaches. The upper bound presents the real experimental results based on our node hardware infrastructure which consumes more energy during sleeping mode (see Section \ref{section:SystemOverview} for details). On the other hand, the lower bound estimates the energy consumption of a hardware infrastructure which supports deep sleep (no consumption during sleeping mode). The need of energy consumption range can be clearly seen in Figures \ref{fig:figTst} and \ref{fig:figTd}. In spite of the sleeping time increase in all cases, the upper bound of energy consumption increases proportionally (the opposite holds for the lower bound). Additionally, PSDETC is expected to consume more energy than the others because of the V-slots. However, Figures \ref{fig:figEst} and \ref{fig:figEd} illustrate the opposite trend for the upper bound.

\subsubsection{\textbf{Effect of ETC Parameters}}
Figure \ref{fig:figEim} presents the effect of different parameters, e.g. $\sigma$, $\rho$, $\mu$ with the same interval length $T=1$.The experiment results follow the trends shown in the theory. In PETC and PSDETC, a smaller $\mathbf{\sigma}$ forces the system to be more conservative and leads to more event condition violations (Figure \ref{fig:figEv}) and consequently to more actuation (Figure \ref{fig:figEact}) and energy consumption (Figure \ref{fig:figEd}). For the same reason, in the decentralized PSDETC, the state transmission reduces with bigger $\mathbf{\sigma}$ (Figure \ref{fig:figEtrans}).

In PADETC, a bigger $\mathbf{\varrho}$ has similar effect as a smaller $\sigma$ in PETC. This can be clearly seen in Figure \ref{fig:figEact} and \ref{fig:figEv}, in which bigger $\mathbf{\varrho}$ causes more actuations and violations respectively. A bigger $\mathbf{\mu}$ can result in more frequent threshold updates, but maintains the threshold less conservative, and thus, the sampling errors can be enlarged. Additionally, Figure \ref{fig:figEv} shows that $\mu$ has greater impact on violations than $\sigma$ and $\varrho$ parameters.

\subsubsection{\textbf{Impact of Interval Length Selection}}
Figure \ref{fig:figTim} illustrates the impact of different interval lengths, in which the same pre-designed Lyapunov converge rate can be guaranteed, for the same set of rest of the parameters, e.g. $\sigma$, $\rho$, $\mu$. %Although a corresponding Lyapunov function can be defined for each $T$ to indicate the convergence of the system, the performance of the system differs.
It can be clearly seen in Figure \ref{fig:figTim} that smaller interval $T$ results in better performance but worse energy consumptions. The water level overshoots are almost the same because of the actuator quantization. Larger sampling times always result in longer convergence time and longer sleeping times. Similarly, the upper bound discharge indicates this trend; longer sleep time leads to higher energy consumption due to the energy hungry operating system background tasks. Oppositely, the lower bound of discharge shows that hardware infrastructures with deep sleep consume significant lower energy for larger interval $T$ due to long sleeping time.%Actuations, valve movement, and violations decrease as the T increases for every approach.
\begin{table*}[ht]
\caption{Savings compared to TTC (\%) total experiment time, i.e. [0, $t_{end}$], with $\sigma=0.2, \rho=85$, and $\mu=0.95$}.
\label{table:savings2}
\centering
%\resizebox{\linewidth}{!}{
\begin{tabular}{c|c|c|c|c|c|c|c|c|}
\hline
\multicolumn{1}{|c|}{\textbf{Approach}} & \textbf{\begin{tabular}[c]{@{}c@{}}Water\\ Level Overshoot\end{tabular}} & \textbf{\begin{tabular}[c]{@{}c@{}}Switching\\   Time\end{tabular}} & \textbf{Discharge} & \textbf{\begin{tabular}[c]{@{}c@{}}Discharge\\   with Deep Sleep\end{tabular}} & \textbf{Actuations} & \textbf{\begin{tabular}[c]{@{}c@{}} Aggr. Valve\\ Movement\end{tabular}} & \textbf{Violations} & \textbf{\begin{tabular}[c]{@{}c@{}}State \\ Transmissions\end{tabular}}\\ \hline
\hline

\multicolumn{1}{|c|}{\textbf{PETC}} & \textbf{3.18} & -0.69 & \textbf{0.97} & \textbf{1.1} & \textbf{18.6} & -1.7 & \textbf{42.8} & 0\\

\hline

\multicolumn{1}{|c|}{\textbf{PSDETC}} & \textbf{1.24} & -0.55 & \textbf{11.03} & -57.6 & \textbf{8.2} & \textbf{1.5} & \textbf{21.5} & \textbf{21.5} \\

\hline

\multicolumn{1}{|c|}{\textbf{PADETC (abs)}} & \textbf{2.44} & \textbf{3.47} & \textbf{4.74} & \textbf{6.9} & \textbf{27.2} & \textbf{7.5} & \textbf{44.8} & \textbf{51.6} \\

\hline

\multicolumn{1}{|c|}{\textbf{PADETC (rel)}} & -19.72 & -20.53 & \textbf{3.28} & \textbf{13.2} & \textbf{9.7} & \textbf{30.5} & \textbf{51.8} & \textbf{56.0} \\

\hline
\end{tabular}
%}
\end{table*}

%\subsubsection{\textbf{Savings Compare to TTC for period from start to switch}}

\begin{table*}[ht]
\caption{Savings compared to TTC (\%) until mode switching time, i.e. [0, $t_{sm}$], with  $\sigma=0.2, \rho=85$, and $\mu=0.95$}.
\label{table:savings}
\centering
%\resizebox{\linewidth}{!}{
\begin{tabular}{c|c|c|c|c|c|c|c|c|}
\hline
\multicolumn{1}{|c|}{\textbf{Approach}} & \textbf{\begin{tabular}[c]{@{}c@{}}Water\\ Level Overshoot\end{tabular}} & \textbf{\begin{tabular}[c]{@{}c@{}}Switching\\   Time\end{tabular}} & \textbf{Discharge} & \textbf{\begin{tabular}[c]{@{}c@{}}Discharge\\   with Deep Sleep\end{tabular}} & \textbf{Actuations} & \textbf{\begin{tabular}[c]{@{}c@{}} Aggr. Valve\\ Movement\end{tabular}} & \textbf{Violations} & \textbf{\begin{tabular}[c]{@{}c@{}}State \\ Transmissions\end{tabular}}\\ \hline
\hline

\multicolumn{1}{|c|}{\textbf{PETC}} & \textbf{3.15} & -0.69 & \textbf{1.34} & \textbf{2.4} & \textbf{30.3} & \textbf{17.3} & \textbf{55.2} & -0.7\\

\hline

\multicolumn{1}{|c|}{\textbf{PSDETC}} & \textbf{1.67} & -0.55 & \textbf{10.97} & -66.9 & \textbf{10.4} & \textbf{8.3} & \textbf{14.6} & \textbf{14.6} \\

\hline

\multicolumn{1}{|c|}{\textbf{PADETC (abs)}} & \textbf{3.02} & \textbf{3.47} & \textbf{9.22} & \textbf{11.9} & \textbf{34.7} & \textbf{24.1} & \textbf{57.0} & \textbf{63.9} \\

\hline

\multicolumn{1}{|c|}{\textbf{PADETC (rel)}} & -19.72 & -20.53 & -15.23 & -4.5 & \textbf{14.4} & \textbf{22.6} & \textbf{49.1} & \textbf{54.5} \\

\hline
\end{tabular}
%}
\end{table*}

\subsubsection{\textbf{Savings Compared to TTC}}
Table \ref{table:savings2} and \ref{table:savings} show the total savings of different ETC techniques against TTC for the time period $period_1=[0, t_{end}]$ (total experiment time) and $period_2=[0, t_{sm}]$ (time until switching mode) respectively. We provide this data separately due to the existence of the switched controller and the different behaviour of the two modes.

In $period_1$ PETC performs similarly to TTC with the difference of  reduced actuations and violations by 18.6\% and 42.8\% respectively. In spite the saving, PETC causes more valve movements than TTC. The PSDETC is more conservative than the centralized PETC, with a result, the lower savings in terms of violations. However, PSDETC reduces the valve movements and the state transmissions due to the decentralized architecture. PADETCabs outperforms all the other approaches because of the asynchronous behaviour, reducing significantly the violations, state transmissions and actuations by achieving 44.8\%, 51.6\%, and 27.2\% savings respectively. PADETCrel occurs similar actuation and communication saving with PADETCrel but with the trade-off of lower performance in terms of water level overshoots and switching time. As has been described in Section \ref{section:ETC}, this happens because the PADETC with reference value updates introduces an extra error, known as maximum dynamic quantization error. However, this extra error allows this triggering mechanism to be more robust against noise than any of the other mechanisms with pre-designed maximum dynamic error.

In $period_2$, some ETC approaches deviate compared to the total savings. For example, in PETC approach, $period_2$ reveals higher actuation savings than $period_1$. The reason is that in mode 1, the weak pump is unable to supply the tanks with enough water and the system deviates from steady state continuously. Thus, event condition violations are being increased and often large valve movements are required. %Another remarkable example is related to the discharge level of PADERCref. In $period_2$, the system consumes around 20\% more energy than $period_1$. This is due to the significantly longer switching time than TTC.
PSDETC and PADETCabs have a more stable behaviour than the other ETC approaches. Again PADETCabs outperforms the other ETC approaches achieving outstanding violation (57\%) and actuation (35\%) savings.

\section{Conclusion}
\label{section:Conclusion}
%In this paper, we have proposed duty-cycling of the sensing and actuator listening activities and enabled decentralized ETC techniques introducing innovative communication schemes. Specifically, we designed and implemented three new MAC layers, which enable the application of four different periodic centralized and decentralized event triggered control approaches. %Additionally, we provided a hybrid controller design and event-triggered control implementation to real plants.
%We provided experimental results, in terms of water level overshoot, energy consumption, sleep time, actuations, valve movement and communication requirements, by implementing one centralized and three decentralized ETC approaches in the extended version of the WaterBox testbed environment \cite{kartakis2015waterbox} and conducting more than 300 experiments. All the approaches provided savings compared to the classic time-triggered control while Periodic Asynchronous Decentralized ETC achieved the best performance. While in this paper we focus on smart water networks, our proposed framework can be applied to a variety of Cyber-Physical Systems such as Smart Grids, Smart Transportation Systems and Automated Agriculture.

In this paper, we have proposed duty-cycling of the sensing and actuator listening activities and enabled decentralized ETC techniques introducing innovative communication schemes. Specifically, we designed and implemented three new MAC layers, which enable the application of four different periodic centralized and decentralized event triggered control approaches. By implementing our proposed communication schemes in the WaterBox testbed \cite{kartakis2015waterbox}, we provided experimental results %, in terms of water level overshoot, energy consumption, sleep time, actuations, valve movement and communication requirements, by implementing one centralized and three decentralized ETC approaches in the extended version of the WaterBox testbed environment \cite{kartakis2015waterbox} and conducting
from more than 300 experiments.

Based on the experimental results, ETC approaches can introduce considerable benefits into industrial deployments. Due to the outstanding decrease of actuations either in number (up to 35\%) or size (i.e. for valve movement up to 24\%), the ETC techniques can increase the robustness, resilience, and lifetime of physical plants and actuators significantly. This increase can lead to significant maintenance cost reduction by postponing expensive replacements of plant assets.

WaterBox consists of energy hungry sensor/ actuator nodes to allow computational intensive algorithm deployments. An optimal hardware infrastructure will reduce the energy consumption even more than the evaluation results. Intuitively, the level of energy reduction will be closer to threshold violations (up to 57\%) and state transmission (up to 64\%) savings which indicate the actuator and sensors communication requirement respectively.

An additional benefit of applying periodic centralized or decentralized ETC approaches is the reduction of sensing rate. Continuous measurement retrieval from high energy demanding sensors (e.g. the water content sensor \cite{wet2016} which consumes 570 mJ per measurement) may lead to higher energy consumption than the communication process (e.g. low power wide area communication modules in \cite{Kartakis2016DLW} which consumes 1.5 to 42 J per 10 bytes). Further, based on our experimental results, higher sensing rates do not guarantee higher control performance. As future work, we will examine the aperiodic sensing scheduling over ETC techniques and the impact to the co-existing high sample rate algorithms for anomaly detection and validation. While in this paper we focus on smart water networks, the proposed framework can be applied to a variety of Cyber-Physical Systems such as Smart Grids, Smart Transportation Systems and Automated Agriculture.

% Our results show that
% \begin{itemize}
% \item ETC approaches can introduce considerable benefits into industrial deployments. Due to the outstanding decrease of actuations (either in number or size), the ETC techniques can increase the robustness, resilience, and lifetime of physical plants and actuators significantly.
% \item Higher sensing rates do not guarantee higher control performance.
% \end{itemize}

%However, WaterBox consists of sensor/ actuator nodes based on the Intel Edison development board, to allow computational intensive algorithm deployments. This lead to high energy consumption. In a real smart water network, a custom made sensor/ actuator node, created based on ETC requirements, will reduce the energy consumption even more than the evaluation results. The  design of this costumed node will be part of our future works. Since measurement retrieval from high energy demanding sensors (e.g. the water content sensor \cite{wet2016} which consumes 570 mJ per measurement) may lead to higher energy consumption than the communication process (e.g. low power wide area communication modules in \cite{Kartakis2016DLW} which consumes 1.5 to 42 J per 10 bytes), examining the aperiodic sensing scheduling over ETC techniques and the impact to the co-existed high sample rate algorithms for anomaly detection and validation will also be part of the future works. While in this paper we focus on smart water networks, our proposed framework can be applied to a variety of Cyber-Physical Systems such as Smart Grids, Smart Transportation Systems and Automated Agriculture.

\section*{Acknowledgment}
\addcontentsline{toc}{section}{Acknowledgment}
This work forms part of the Big Data Technology for Smart Water Networks research project funded by NEC Corp, Japan and partly supported by China Scholarship Council (CSC).

\bibliographystyle{IEEEtran}

\end{document}